\documentclass[preprint]{elsarticle}

\usepackage{graphicx}
\usepackage{booktabs}
\usepackage{multirow}
\usepackage{algorithm}
\usepackage{breqn} 
\usepackage{algpseudocode}
\usepackage{xcolor}
\usepackage{amsmath}
\usepackage{amssymb}
\usepackage{amsthm}
\usepackage{subcaption}
\usepackage{mathtools}
\usepackage{multicol}
\usepackage{bm}
\usepackage{url}

\algnewcommand\algorithmicassert{\texttt{assert}}
\algnewcommand\algdelete{\texttt{delete}}
\algnewcommand\Assert[1]{\State \algorithmicassert #1}
\algnewcommand\Delete[1]{\State \algdelete #1}

\newtheorem{theorem}{Theorem}
\theoremstyle{definition}
\newtheorem{definition}{Definition}[section]

\mathchardef\mhyphen="2D

\begin{document}

\title{Attribute-Based Authentication in Secure Group Messaging for Distributed Environments and Safer Online Spaces}

\author[1]{David Soler \corref{cor1}}
\ead{david.soler@udc.es}
\author[1]{Carlos Dafonte}
\author[2]{Manuel Fernández-Veiga}
\author[2]{Ana Fernández Vilas}
\author[1]{Francisco J. Nóvoa}

\cortext[cor1]{Corresponding author}
\affiliation[1]{organization={CITIC, Universidade da Coruña},
            city={A Coruña},
            country={Spain}}
\affiliation[2]{organization={atlanTTic, Universidade de Vigo},
            city={Vigo},
            country={Spain}}

\begin{abstract}
The Messaging Layer security (MLS) and its underlying Continuous Group Key Agreement (CGKA) protocol allows a group of users to share a cryptographic secret in a dynamic manner, such that the secret is modified in member insertions and deletions. Although this flexibility makes MLS ideal for implementations in distributed environments, a number of issues need to be overcome. Particularly, the use of digital certificates for authentication in a group goes against the group members' privacy. In this work we provide an alternative method of authentication in which the solicitors, instead of revealing their identity, only need to prove possession of certain attributes, dynamically defined by the group, to become a member. Instead of digital certificates, we employ \textit{Attribute-Based Credentials} accompanied with Selective Disclosure in order to reveal the minimum required amount of information and to prevent attackers from linking the activity of a user through multiple groups. We formally define a CGKA variant named \textit{Attribute-Authenticated Continuous Group Key Agreement} (\textit{AA-CGKA}) and provide security proofs for its properties of Requirement Integrity, Unforgeability and Unlinkability. We also provide an implementation of our AA-CGKA scheme and show that it achieves performance similar to a trivial certificate-based solution.
\end{abstract}

\begin{keyword}
CGKA \sep MLS \sep Attribute-Based Credentials
\end{keyword}

\maketitle

\section{Introduction} 
\label{sec:introduction}

Messaging Layer Security (MLS)~\cite{mls} is a recent communications standard for establishing secure messaging groups between a set of users. It is mainly composed by a cryptographic scheme named Continuous Group Key Agreement (CGKA) \cite{modular}, whose objective is to distribute a shared secret between members of the group. One of the most relevant characteristics of CGKA protocols is its flexibility in group composition: the scheme allows insertions and deletions of members in an efficient manner, while ensuring the security properties of Forward Secrecy (FS) and Post-Compromise Security (PCS). 

The dynamic nature of CGKA protocols makes them interesting for distributed environments. These networks are defined by their lack of centralised agents in which users implicitly trust, whether it be for providing authentication, forwarding information between parties or establishing items for common reference. Privacy is also a primary concern in distributed environments, as there are no pre-existing trust relationships between users: they wish to reveal as little information as possible about their identity or other activities they perform in the network. CGKA protocols could be employed in distributed environments to establish communication between parties without prior knowledge of each other but nonetheless share common attributes or interests. 

However, there are a number of problems that prevent CGKA schemes from being applicable to a distributed setting. The authentication method assumed by these CGKA schemes (and made explicit in the MLS specification) relies on the use of long-term signature keys associated to an user's identity through a digital certificate. This implies that all authentication attempts performed by the same user employ the same key material. Thus, it is possible to track the activity of users across all groups they participate in. Furthermore, a verifier must have access to the user's digital certificate in order to verify their long-term signature key. This would reveal the user's full identity, which goes against the user's privacy.

In this work we develop an alternative CGKA scheme suitable for its implementation for distributed environments, in which the aforementioned problems are addressed. Instead of validating the identity of new users, the group will check if they fulfil certain \textit{requirements} ---such as possession of a university degree or being above certain age--- that can be dynamically modified by group members. This way, users aiming to join the group would only need to reveal the required information while keeping their identity in private. We also remove the need of long-term signature keys: users are able to generate fresh key pairs for every group which are nevertheless verifiable by existing group members.

To that end, we replace digital certificates with Attribute-Based Credentials (ABC) \cite{vc_20, ssi_survey}. Instead of being linked to a real identity, these kind of credentials certify that its owner is in possession of specific attributes, called \textit{claims}. Attribute-Based Credential schemes usually allow holders of credentials to perform \textit{Selective Disclosure} such that only a subset of the attributes are revealed while remaining a valid credential. Furthermore, the \textit{Unlinkability} property of ABC schemes ensures that it is impossible to know if two different authentication attempts have been generated from the same credential. 

Our resulting scheme is named \textit{Attribute-Authenticated CGKA} (\textit{AA-CGKA}). We provide a formal specification of our AA-CGKA scheme and define the security properties of Requirement Integrity, Unforgeability and Unlinkability. We then perform a formal security analysis proving that our construction meets said definitions. 

We also present an implementation of our AA-CGKA scheme in MLS to demonstrate its applicability to real world scenarios. We make use of Extensions, a mechanism defined by MLS to modify its functionality, to include the set of requirements into the group's state and to define new proposal types for modifying it. We envision our AA-CGKA scheme to be useful in healthcare environments and in the protection of online communities populated by minors, among other applications.

In summary, our work presents the following contributions:
\begin{enumerate}
    \item The AA-CGKA scheme, in which users are able to join a group only if they possess a set of specific claims, dynamically defined by the group in the form of requirements. Solicitors employ Attribute-Based Credentials to prove possession of the group's claims, which allows them to use Selective Disclosure and protects them from being tracked across multiple AA-CGKA groups. We construct our AA-CGKA protocol using as primitives a CGKA protocol and an ABC scheme, and we provide a detailed formal specification in which we define the behaviour of our proposed protocol. 
    \item A security analysis for our proposed AA-CGKA scheme. We define the security properties of Requirement Integrity, Unforgeability and Unlinkability. Informally, these properties ensure that only users that fulfil the group's requirements can join an AA-CGKA group, and that it is impossible to obtain any private information about a solicitor by their request to join. Then, we perform a formal analysis of our proposal proving that it fulfils all three security notions. This represents an improvement over other CGKA variants, as their requirement to employ long-term signature keys ensures that the activities of users can be linked across different groups.
    \item An implementation of our AA-CGKA scheme into MLS. We employ MLS' extensibility options to define new types of credentials, proposals and data structures. We show that our scheme introduces little overhead in commit generation and processing times compared to a traditional certificate-based solution.
\end{enumerate}

The rest of the document is organised as follows: Section~\ref{sec:literature} will review related works to provide the context to our contribution. Section~\ref{sec:tech} will introduce the required technical background to the reader, including the cryptographic primitives we will employ to define our scheme. In Section~\ref{sec:protocol} we will provide a formal specification of our proposal, as well as a general overview of our target scenario. Section~\ref{sec:security} discusses the desired security properties of the protocol and provides formal security proofs. Section~\ref{sec:impl} introduces the implementation of our AA-CGKA scheme and its experimental measurements. In Section~\ref{sec:discussion} we analyse the obtained results and discuss practical considerations related to a real-world deployment of our scheme. Finally, Section~\ref{sec:conclusion} will conclude this document and discuss future improvements.

\section{Related Work}
\label{sec:literature}

CGKA protocols have been thoroughly defined in the literature~\cite{cgka_analysis, itk}, exploring the security properties of Forward Security (FS) and Post-Compromise Security (PCS)~\cite{cgka_analysis}. These schemes usually aim for logarithmic complexity in insertions and removals \cite{bounds, art, treekem}. The most popular instantiation of a CGKA protocol is the Messaging Layer Security (MLS), which has recently been standardised as RFC 9420 \cite{mls}. MLS employs TreeKEM \cite{treekem} as its CGKA protocol, and uses the shared cryptographic secret to perform Secure Group Messaging between its members.  

The dynamic nature of CGKA protocols make them particularly useful for distributed settings. However, there are numerous challenges that need to be overcome, specially due to the possibility of concurrent and incompatible updates to the state of the group. The authors of \cite{fork} develop a method for resolving forks while maintaining Forward Secrecy. In \cite{dec_ack}, the authors present a scheme in which the order in which concurrent updates are processed is irrelevant, but does not achieve logarithmic complexity for updates. The protocol presented in~\cite{decaf} is a refinement of the Server-Aided CGKA of \cite{saik}, in which the server is replaced with a blockchain and multiple updates are applied in bulk. However, these works either reduce the efficiency of the protocol or require a relaxed notion of FS or PCS. 

Other works aim to provide variants to CGKA protocols to achieve specific objectives unrelated to efficiency or its security guarantees. The \textit{Administrated CGKA} protocol of \cite{a_cgka} divides the group in administrators and non-admins, such that only the former members can issue certain updates. In the protocol presented in \cite{cgka_anon} messages are anonymous such that other members of the group cannot know who is publishing each message.  


Attribute-Based Credential schemes \cite{ssi_survey, issuer_hiding} provide an alternative to traditional certificates by containing a set of \textit{verifiable claims} about its holder. Selective Disclosure techniques \cite{sd_jwt_draft, bbs} allow holders to only show a subset of a credential's claims, thus increasing privacy. The Verifiable Credentials data model \cite{vc_20}, developed by W3C, is a popular instantiation of an ABC scheme. Attribute-Based Signature schemes \cite{abs-original}, particularly Unlinkable variants \cite{abs-1, abs-2}, provide similar functionality; however, they are less common and thus less compatible with existing identity infrastructure.

The use of Attribute-Based Credentials in MLS is introduced in the draft \cite{mls_vc_draft}, which was uploaded concurrently to our work. However, the focus of the draft is substantially different than ours: the Attribute-Based Credentials are treated as if they were X.509 certificates, employed only to verify the solicitor's identity.

\section{Background} 
\label{sec:tech}

\subsection{Continuous Group Key Agreement}
\label{sec:cgka}

A CGKA protocol is a scheme that allows a set of users to establish a common secret. This shared value changes dynamically with every modification of the state of the group, whether it be insertions or eliminations of users. Each member possesses some secret information which they are allowed to update if they believe it has been compromised. Each state of the group is called an \textit{epoch}; whenever a member issues a modification to the state of the group through a \textit{Commit}, a new epoch is created with a different shared secret.



Formally, a CGKA scheme is composed of a tuple of algorithms ($\mathsf{Init}$, $\mathsf{Create}$,$\mathsf{GenJP}$, $\mathsf{Propose}$, $\mathsf{Commit}$, $\mathsf{Process}$), defined as follows:

\begin{enumerate}
    \item $\gamma \gets \mathsf{Init}(1^\lambda, id)$: Initialises $id$'s state. Takes as parameters the security parameter $1^\lambda$ the user's ID $id$. 
    \item $\gamma' \gets \mathsf{Create}(\gamma)$: Creates a new CGKA group with the initiator as its only member. Takes as parameter the initiator's state $\gamma$. 
    \item $JP \gets \mathsf{GenJP}(\gamma, jp\mhyphen type)$: Generates a \textit{Join Package} from the user's state $\gamma$. Also takes as parameter the Join Package type $jp\mhyphen type$, which is either $\mathsf{add}$ or $\mathsf{external}$. 
    \item $(\gamma', prop) \gets \mathsf{Propose}(\gamma, id, prop\mhyphen type, JP)$: Creates a new proposal of type $prop$-$type$, whose possible values are $\{\mathsf{add}, \mathsf{update}, \mathsf{remove}, \mathsf{external}\}$. Takes as parameters the proposer's state $\gamma$ and the ID of the affected user $id$. Additionally, $\mathsf{add}$ and $\mathsf{external}$ proposals also take as parameter the new user's Join Package $JP$. 
    \item $(\gamma', C) \gets \mathsf{Commit}(\gamma, p_L = [prop_1, \dots, prop_n])$: Commits a list of proposals $p_L$ and outputs a commit message $C$. Takes as parameter the committer's state $\gamma$.
    \item $(\gamma', ok) \gets \mathsf{Process}(\gamma, C)$: Processes a commit message $C$ generated by another user. If $C$ is validated correctly (indicated by the Boolean value $ok$), the user's state $\gamma$ is updated.
\end{enumerate}

We include in the definition of a CGKA protocol the proposal type $\mathsf{external}$, which encapsulates the functionality of \textit{External Joins}. Using this operation, users who are not part of a CGKA group are able to directly enter one without requiring an $\mathsf{add}$ proposal from a current member. The creator of the $\mathsf{external}$ proposal can commit it and send it to the members without being part of a CGKA group. This eliminates the need for invitation to participate in CGKA groups, providing a more agile method for joining them. External Joins are defined in the MLS RFC~\cite{mls} and have only recently been included in formal security analysis of CGKA schemes \cite{cgka_analysis_etk}.


The security of a CGKA protocol is usually established in a \textit{Key Indistinguishability} attack game \cite{a_cgka, cgka_analysis}, in which it is proven that an attacker cannot obtain any information about the secret key of a CGKA group in an specific epoch, even with access to all control messages generated through the evolution of the group. A more detailed definition of this security property is located in \ref{app:cgka}.

We define a structure called \textit{Join Package} as a parameter for proposals, which encapsulates some cryptographic material required to insert an user into a CGKA group. We also explicitly define the algorithm $\mathsf{GenJP}$ for creating Join Packages. Readers familiar with MLS terminology should interpret Join Packages as a combination of KeyPackages for $\mathsf{add}$ proposals and ExternalInit for $\mathsf{external}$ proposals \cite{cgka_analysis_etk}. We bundle these two structures into one for convenience, as will become apparent when we define our AA-CGKA scheme. The contents of Join Packages are determined by their type $jp\mhyphen type$:

\begin{enumerate}
    \item The $\mathsf{add}$ Join Packages contain cryptographic information that ensures that only the creator of the Join Package can decrypt the \textit{Welcome message} that is generated when the proposal is committed. 
    \item The $\mathsf{external}$ Join Packages contain a key $k$ that is employed to derive the new epoch's shared secret. This is required because external joiners do not have information about the current state of the group. Only the current members of the group (including the joiner) can obtain $k$. 
\end{enumerate}

While we are agnostic to the structure of a Join Package, we require that the above properties hold. Clearly, this is required for the Key Indistinguishability security property: otherwise, an attacker could obtain the shared secret by decrypting the \textit{Welcome message} (in the case of $\mathsf{add}$ Join Packages) or $k$ (for $\mathsf{external}$ Join Packages). 

\subsection{Attribute-Based Credentials}
\label{sec:abc}

Attribute-Based Credentials (\textit{ABC}), also known as Anonymous Credentials \cite{issuer_hiding, urs, abc_fw} are a type of personal document that contain a set of claims about its holder. These claims correspond to attributes holders possess: for example, they could contain their age or possession of a certain academic degree.

They differ from traditional X.509 digital certificates in that ABCs are focused toward the holder's privacy: the credentials only contain the set of claims required for authentication instead of the holder's identity, which can be employed to link their activity across multiple services. For that reason, ABCs are employed in modern Identity Management paradigms such as Federated or Self-Sovereign Identity models \cite{ssi_survey}. 

\paragraph*{\textbf{Selective Disclosure}}

ABC schemes allow holders to only reveal a certain subset of attributes from their credentials, such that their privacy is maintained in case only some of their attributes are required for authentication. This means either removing some claims from the credential or transforming them: for example, it may be useful to disclose only that the holder's age is above 18 without revealing the exact value \cite{issuer_hiding}.

For this reason, we distinguish two types of documents in an ABC scheme: credentials and Presentations. The latter is generated from the holder from a credential, does not require interaction with the issuer \cite{vc_20} and contains only a subset of its claims. In order to be validated, Presentations contain a \textit{Proof of Knowledge} of the credential's signature \cite{cl}. 

Formally, an ABC scheme is composed of the following operations~\cite{bbs, urs}:

\begin{enumerate}
    \item $(isk, ipk) \gets \mathsf{KeyGen}(1^\lambda)$: Initialises the public and private keys for the issuer. Takes as parameter the security parameter $1^\lambda$.
    \item $cred \gets \mathsf{Issue}(isk, attrs)$: Issues a credential $cred = (attrs, \sigma)$ signed with the issuer's secret key $isk$, which certifies possession of a set of attributes $attr$. The credential contains a signature $\sigma$.
    \item $b \gets \mathsf{VerifyCred}(ipk, cred)$: Verifies the credential $cred$ using the issuer's public key $ipk$. Outputs $true$ or $false$ depending on the validity of $cred$.
    \item $P/\bot \gets \mathsf{Prove}(ipk, cred, attrs, header)$: Generates a Presentation $P = (header, attrs, \pi)$ which contains a list of disclosed attributes $attrs$, a header $header$ and a zero-knowledge proof $\pi$ which proves knowledge of the signature $\sigma$ contained in $cred$. Also takes as parameter the issuer's public key $ipk$ and a header $header$. The algorithm fails if $attrs$ are not contained in $cred$.
    \item $b \gets \mathsf{VerifyProof}(ipk, P, attrs, header)$: Verifies the Presentation $P$ with the issuer's public key $ipk$. Also takes as parameter the $header$ and target attributes $attrs$, and outputs $true$ or $false$ depending on the validity of $P$'s proof $\pi$.
\end{enumerate}

The following security properties are defined for an ABC scheme \cite{issuer_hiding}:
\begin{itemize}
    \item Unforgeability: it is impossible to generate a valid Presentation $P$ proving possession of attributes $disc\mhyphen attrs$ without access to a credential $cred$ that contains at least $disc\mhyphen attrs$.
    \item Unlinkability: given credentials $cred_0$ and $cred_1$ and Presentation $P_b$ generated from one of the credentials as $P_b \gets \mathsf{Prove}(ipk, cred_b, disc\mhyphen attrs, header)$, the probability for an attacker to calculate $b$ is no better than a random guess (as long as both $cred_0$ and $cred_1$ contain the attributes of $disc\mhyphen attrs$). This property holds even if the attacker controls the issuer and is able to define $disc\mhyphen attrs, header$ and $(cred_0, cred_1)$.
\end{itemize}

As with the CGKA scheme, we include more detailed definitions of the security properties of Unforgeability and Unlinkability in \ref{app:abc}.

\section{Attribute-Authenticated Continuous Group Key Agreement}
\label{sec:protocol}

In this Section, we will introduce our proposed AA-CGKA protocol. We will start by introducing the main entities that take part in the execution of our protocol. Then, we will provide a high-level overview of the protocol in which we will describe the main operations involved in its execution. Finally, we will introduce the formal specification of our protocol, defining the functions it is composed of and their inputs and outputs.

\subsection{Outline}
\label{sec:outline}

The main entities that take part in our AA-CGKA protocol are the following:

\begin{enumerate}
    \item \textit{Credential Issuer}: third party who creates credentials on behalf of users. Their signature grants validity to the credentials they issue. 
    \item \textit{Solicitor}: user who is trying to join an AA-CGKA group. To this end, the solicitor must obtain the Group Information Message, which contains relevant information about the group and then issue an External Join to directly enter the group without requiring invitation from an existing member. The solicitor must also include a Presentation in which she includes a list of claims that meet the group's requirements. Solicitors store their credentials in a \textit{wallet}, which also automatically generates Presentations in which only certain attributes are disclosed.
    \item \textit{Verifiers}: current members of the AA-CGKA group. When a group is initialised the only member is its creator, but new users eventually join. Every current member of the AA-CGKA group can propose modifications to the group including changes to the set of requirements, and publish a Group Information message. 
\end{enumerate}

\begin{figure}[t]
    \centering
    \includegraphics[width=0.9\linewidth]{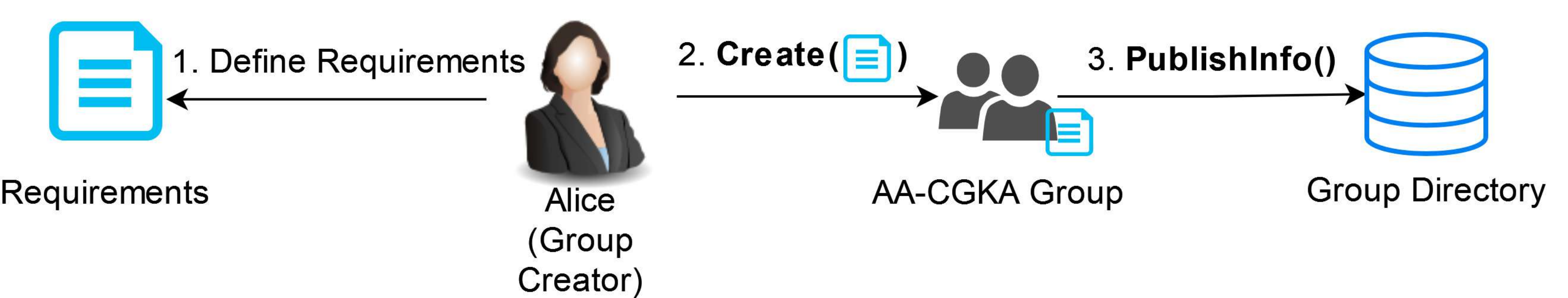}
    \caption{Overview of the $\mathsf{Create}$ and $\mathsf{PublishInfo}$ operations.}
    \label{fig:create}
\end{figure}

Figure~\ref{fig:create} shows the procedure by which an AA-CGKA group is created. The group creator Alice must first generate a set of requirements specifying which claims all solicitors must meet. Initially, Alice is the only member of the group and thus the only \textit{verifier} for future solicitors. 

\paragraph*{\textbf{Requirements}}

\begin{figure}
    \centering
    \includegraphics[width=0.8\linewidth]{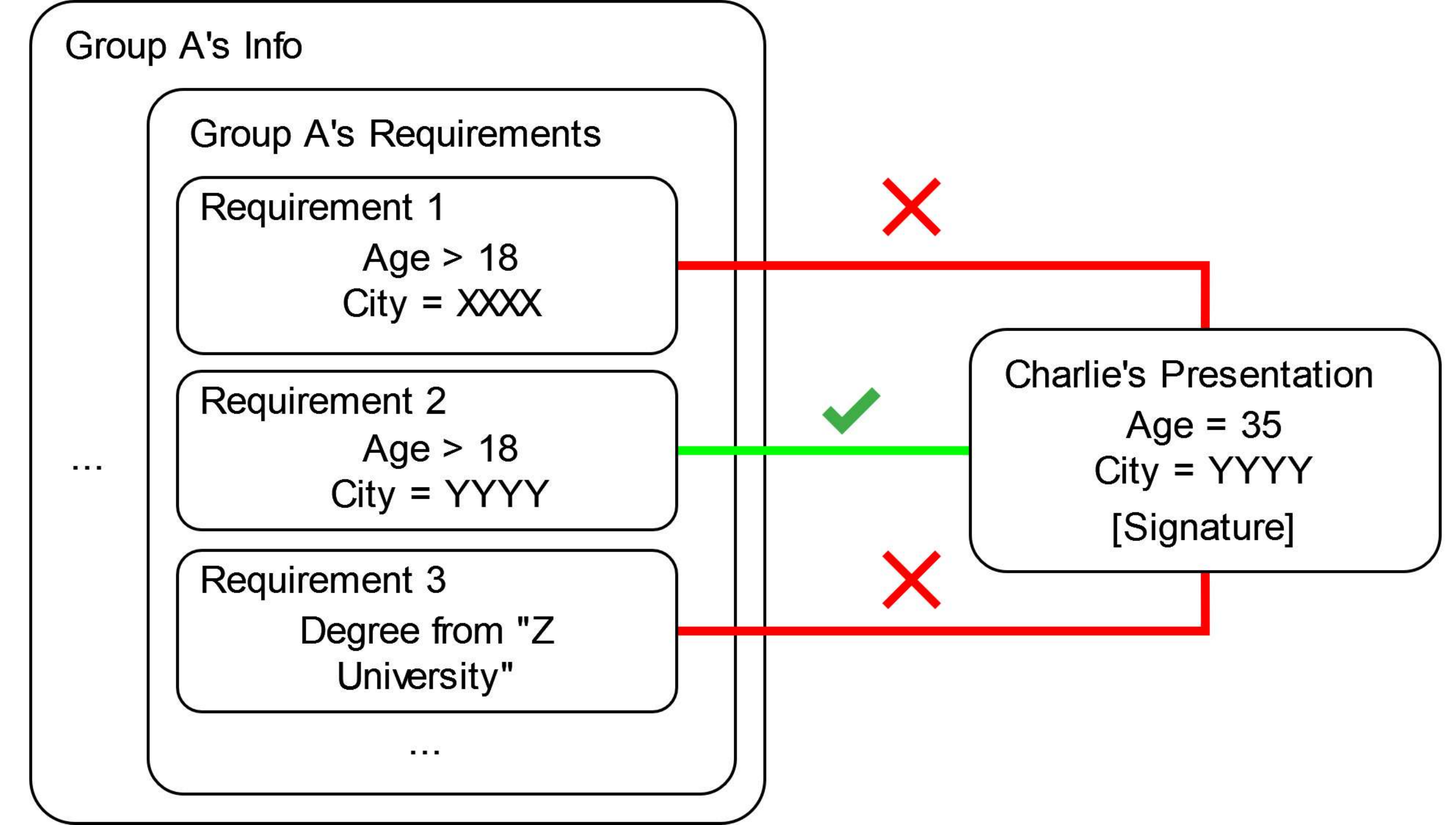}
    \caption{Example of a set of requirements.}
    \label{fig:reqs}
\end{figure}

In our AA-CGKA protocol, each group can establish a set of requirements that all solicitors must meet. Specifically, each of these requirements specifies a list of claims in the form of a key-value pair. The solicitor's Presentation must match all claims in some of the group's requirements. This allows for flexibility in the definition of the requirements, such that a group can specify different profiles of users who can join the group. Figure~\ref{fig:reqs} shows an example in which an AA-CGKA group publishes a Group Info message with a set of 3 different requirements. 

\paragraph*{\textbf{Joining a group}}

As with a standard CGKA scheme \cite{cgka_analysis_etk}, we provide two methods for becoming a member of an AA-CGKA group: either through an invitation from an external member (by generating and $\mathsf{add}$ proposal) or by performing an \textit{External Join}. This latter option allows users to enter a group without being explicitly invited by someone who is already inside. We believe that this method for accessing groups is more suited for a distributed setting, because it allows for group compositions in which users have no prior relationships to each other. 

Users attempting to join an AA-CGKA group must create a Presentation Package. This data structure is composed of a Join Package (as required by the CGKA scheme) but also includes an ABC Presentation containing the solicitor's attributes. Solicitors generate a Presentation Package using the AA-CGKA function $\mathsf{Present}$. The creation of a Presentation Package requires obtaining some information about the state of the group, including the current set of requirements. This information must be generated by a group member through the function $\mathsf{PublishInfo}$ and made publicly available, as shown in Figure \ref{fig:create}.

\begin{figure}[t]
    \centering
    \includegraphics[width=0.9\linewidth]{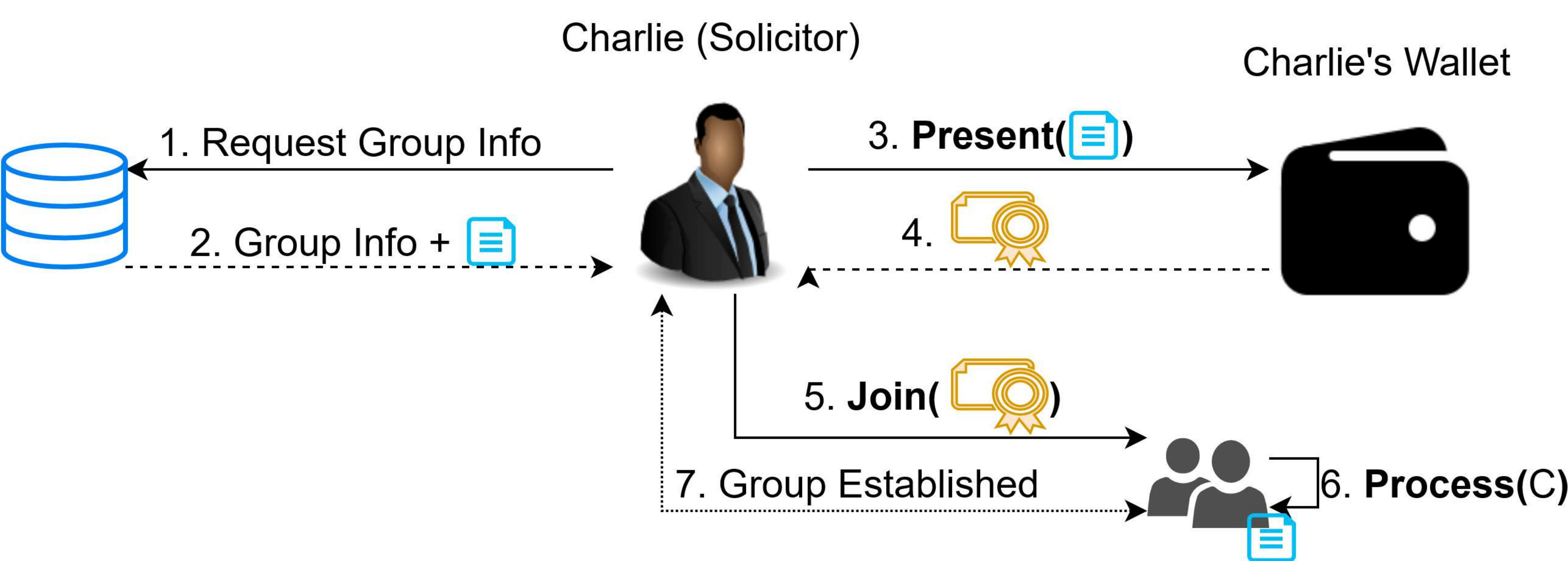}
    \caption{Overview of the $\mathsf{Present}$ and \textit{External Join} operations.}
    \label{fig:join}
\end{figure}

The process of joining a group is shown in Figure~\ref{fig:join}. The solicitor Charlie first obtains the AA-CGKA group information from a public repository, as was published in Figure~\ref{fig:create}. Then, Charlie checks if he meets any of the requirements through his wallet, in which he holds his ABCs. If successful, he generates a Presentation Package which contains a proof that he meets the required claims, as well as other information (such as cryptographic keys) that is required by the underlying CGKA protocol to join a group. Upon receiving the Presentation Package, the AA-CGKA group members process the message and, if it is successfully validated, Charlie becomes a member.

\paragraph*{\textbf{Signature Keys}}

In traditional CGKA protocols, it is assumed that a Public Key Infrastructure (PKI) exists outside the protocol~\cite{cgka_analysis, a_cgka} in which long-term signature keys are associated to any user's identity. 

However, our model is slightly different since the ABCs that solicitors present do not always carry long-term key material. This is intentional: key material reuse would imply that a verifier could identify the same solicitor in different CGKA groups or other online activities, which would work in detriment to the solicitor's privacy. To avoid this, we require that solicitors create a fresh signature key pair and send it to the CGKA group inside the Presentation. We remark that this signature key pair is different from the Hybrid Public Key Encryption (HPKE) key pair that is usually included in a Join Package as specified by CGKA protocols~\cite{mls}, which is also a fresh key pair generated to join a CGKA group but is employed for receiving the group's relevant secrets and not for message authentication. 

This signature key pair needs to be associated to the user's credential nonetheless. If they were not, an attacker could simply take another user's Presentation, generate a new signature key pair and present it as its own to the verifiers. To avoid this, we require that the signature public key is included in the header of the Presentation when executing the ABC operation $\mathsf{Prove}$, such that the key pair is linked to this user's Presentation but different keys can be used in multiple Presentations of the same credential.


\paragraph*{\textbf{Modifications to the group}}

\begin{figure}[t]
    \centering
    \includegraphics[width=0.9\linewidth]{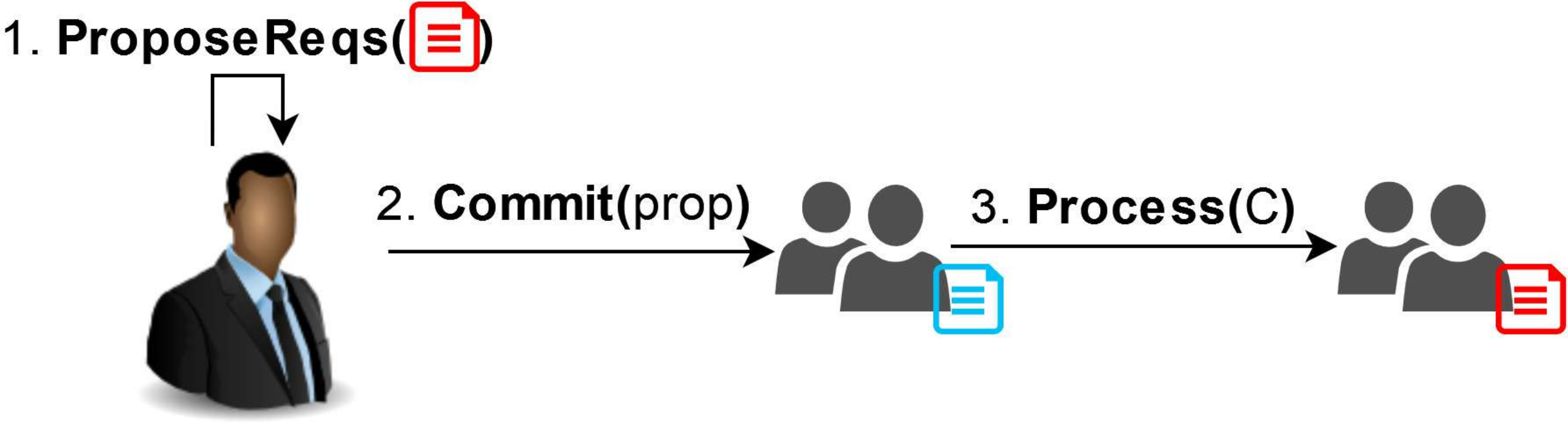}
    \caption{Overview of the $\mathsf{ProposeReqs}$, $\mathsf{Commit}$ and $\mathsf{Process}$ operations.}
    \label{fig:update}
\end{figure}

Our AA-CGKA protocol uses the \textit{propose-and-commit} paradigm of CGKA schemes not only for management of current members, but also to modify their group's set of requirements. To that end we include a new algorithm $\mathsf{ProposeReqs}$. The procedure for this is shown in Figure~\ref{fig:update}: at any moment, a member of the AA-CGKA group can issue a proposal to modify the current requirements of the group. This proposal is then committed by the same or a different member. 

The commit is processed by the rest of group members through the $\mathsf{Process}$ algorithm, modifying the set of requirements to enter the group. This modification may consist in an insertion of a new requirement or the update or deletion of an existing one. We remark that this method of processing updates is inherited by our protocol from standard CGKA definitions: methods for avoiding or resolving conflicts among alternative proposals or commits are outside the scope of this work \cite{decaf, fork}. 

\subsection{Formal specification}

An \textit{Attribute-Authenticated CGKA} (AA-CGKA) scheme is composed of a tuple of algorithms ($\mathsf{Init}$, $\mathsf{Create}$, $\mathsf{Propose}$, $\mathsf{ProposeReqs}$, $\mathsf{Commit}$, $\mathsf{Process}$, $\mathsf{PublishInfo}$, $\mathsf{Present}$), defined as follows:

\begin{enumerate}
    \item $\mathsf{Commit}$ and $\mathsf{Process}$ are defined as they were for a CGKA.
    \item $\gamma \gets \mathsf{Init}(1^\lambda, ID, attrs, issuer)$ additionally takes as parameters attributes \emph{attrs} and issuer $issuer$.
    \item $\mathsf{Create}(\gamma, reqs)$ additionally takes as parameter a set of requirements $reqs$. 
    \item $\mathsf{Propose}(\gamma, id, prop\mhyphen type, PP)$ is modified to take as parameter a Presentation Package $PP$ instead of a Join Package.
    \item $\mathsf{ProposeReqs}(\gamma, prop\mhyphen type, i, new\mhyphen req)$: adds new requirements to the group, or modifies or deletes existing ones. The operation to perform is determined by $prop\mhyphen type \in \{\mathsf{add}, \mathsf{update}, \mathsf{remove}\}$, defined as for $\mathsf{Propose}$ but without $\mathsf{external}$. Also takes as parameter the index of an existing requirement $i$ and a new requirement $new$-$req$.
    \item $GI \gets \mathsf{PublishInfo}(\gamma)$: Creates a Group Information message $GI$ required to enter the group through external joins. 
    \item $PP \gets \mathsf{Present}(1^\lambda, \gamma, GI)$: Generates a Presentation Package $PP$ to join a CGKA group, using the group information $GI$. Also takes as parameters a User state $\gamma$, and outputs a Presentation Package $PP$. 
\end{enumerate}

An AA-CGKA protocol uses as primitives a CGKA scheme \textit{CGKA = (CGKA.Init, CGKA.Create, CGKA.GenJP, CGKA.Propose, CGKA.Commit, CGKA.Process)}, an ABC scheme \textit{ABC = (ABC.KeyGen, ABC.Issue, ABC.VerifyCred, ABC.Prove, ABC.VerifyProof)}, an EUF-CMA signature scheme \textit{S = (S.GenKeyPair, S.Sign, S.Verify)} and a hash function $H$. The security requirements for these primitives are detailed in \ref{app:games}.

\subsection{Construction}

We introduce a concrete construction of our protocol in Figure \ref{alg:def}, with auxiliary functions defined in Figure \ref{alg:def-aux}. We also include a simplified model of a PKI in Figure \ref{alg:def-pki}, which is tasked with storing the issuer key pairs and providing the public keys to any user that requests them.

\begin{table}
\centering
\caption{List of notations of the proposed scheme.}
\begin{tabular}{ll}
\toprule
Name     & Description                         \\ 
\midrule
$(spk, ssk)$    & Signature Key Pair \\
$chal$ & Challenge bytes             \\
$cred$  & User Attribute-Based Credential              \\
$attrs$ & User Attributes \\
$reqs$ & Set of requirements of the group       \\
$state$      & Base CGKA state              \\
$prop_{TBS}$    & Proposal (To be signed)     \\
$JP_{TBS}$ & Join Package (To be signed)     \\
$C_{basic}$ & Committed Base CGKA Proposals \\
$C_{reqs}$ & Committed Requirement proposals \\
$W$   & \textit{Welcome} message             \\ 
\bottomrule
\end{tabular}
\label{table:notation}
\end{table}

\begin{figure*}
\begin{multicols}{2}
\tiny
\begin{algorithmic}

\Function{Init}{$1^\lambda, id, attrs, issuer$}
\State $\gamma.self \gets id$
\State $\gamma.sec \gets 1^\lambda$
\State $\gamma.attrs \gets attrs$
\State $\gamma.issuer \gets issuer$
\State $(\gamma.ipk, \gamma.cred) \gets issueCredential(attrs, issuer)$
\State $\gamma.state \gets \mathsf{CGKA.Init}(1^\lambda, \gamma.self)$
\State \Return{$\gamma$}
\EndFunction

\\

\Function{Create}{$\gamma, reqs$}
\State $(\gamma.state.spk, \gamma.state.ssk) \gets \mathsf{S.GenKeyPair}(\gamma.sec)$
\State $\gamma.chal \gets ""$
\State $\gamma.reqs \gets reqs$
\State $\gamma.state \gets \mathsf{CGKA.Create}(\gamma.state)$
\EndFunction

\\

\Function{Propose}{$\gamma, id, prop\mhyphen type, PP$}
    \If{$prop\mhyphen type = \mathsf{add} \lor prop\mhyphen type = \mathsf{external}$}
        \State $prop \gets createAddProposal(\gamma, id, PP)$
        \State \Return{$prop$}
    \Else
        \State $prop \gets \mathsf{CGKA.Propose}(\gamma, id, prop\mhyphen type, \_)$
    \EndIf
    \State \Return{$prop$}
\EndFunction

\\

\Function{ProposeReqs}{$\gamma, prop\mhyphen type, id, new\mhyphen reqs$}
    \If{$prop\mhyphen type = \mathsf{add}$}
        \State $prop_{TBS} \gets (\gamma.self, prop\mhyphen type, new\mhyphen reqs)$
    \ElsIf{$prop\mhyphen type = \mathsf{update}$}
        \State $prop_{TBS} \gets (\gamma.self, prop\mhyphen type, id, new\mhyphen reqs)$
    \ElsIf{$prop\mhyphen type = \mathsf{remove}$}
        \State $prop_{TBS} \gets (\gamma.self, prop\mhyphen type, id)$
    \EndIf
    \State $prop \gets (prop_{TBS}, \mathsf{S.Sign}(\gamma.state.ssk, prop_{TBS}))$
    \State \Return{$prop$}
\EndFunction

\\

\Function{Commit}{$\gamma, P_L$}
    \State $(basic\mhyphen props, req\mhyphen props) \gets organiseProps(P_L)$
    \Assert{ $validateProps(\gamma, basic\mhyphen props, req\mhyphen props)$}
    \State $C_{basic} \gets \mathsf{CGKA.Commit}(\gamma.state, basic\mhyphen props)$
    \If {$C.welcome$ exists}
        \State $W \gets (C.welcome, \gamma.self, \gamma.reqs)$
    \EndIf
    \State $C_{reqs} = req\mhyphen props$
    \State $\sigma \gets \mathsf{S.Sign}(\gamma.state.ssk, (\gamma.self, C_{basic}, C_{reqs}, W, \gamma.chal))$
    \State \Return{$(\gamma.self, C_{basic}, C_{reqs}, W, \sigma)$}
\EndFunction

\\

\Function{Process}{$\gamma, C$}
    \State $(id, C_{basic}, C_{reqs}, W, \sigma) \gets C$
    \If{$\gamma.state.members[id]$ exists}
        \State $pk \gets \gamma.state.members[id].spk$
    \Else
        \State $PP_{joiner} \gets C_{basic}.props[0].PP$
        \State $pk \gets PP_{joiner}.spk$
    \EndIf
    \If{$\lnot \mathsf{S.Verify}(pk, (id, C_{basic}, C_{reqs}, W, \gamma.chal), \sigma)$} \State \Return{$false$} \EndIf
    \If {$\gamma.self \notin \gamma.state.members \land W exists$}
        \State $\gamma.reqs \gets W.reqs$
        \State $(\gamma.state, ok) \gets \mathsf{CGKA.Process}(\gamma.state, W.welcome)$
    \Else
        \State $basic\mhyphen props \gets C_{basic}.props$
        \If{$\lnot validateProps(\gamma, basic\mhyphen props, C_{reqs})$} \State \Return{$false$} \EndIf
        \State $(\gamma.state, ok) \gets \mathsf{CGKA.Process}(\gamma.state, C_{basic})$
        \If{$\lnot ok$} \State \Return{$false$} \EndIf
        \State $\gamma.reqs \gets updateReqs(\gamma.reqs, req\mhyphen props)$
        \State $\gamma.chal \gets updateGroupChal(\gamma.chal, C)$
    \EndIf
    \State \Return{$ok$}
\EndFunction

\\

\Function{PublishInfo}{$\gamma$}
    \State $chal \gets \gamma.chal$
    \State $reqs \gets \gamma.reqs$
    \State $GI = (chal, reqs)$
    \State \Return{$GI$}
\EndFunction

\\

\Function{Present}{$\gamma, GI, jp\mhyphen type$}
    \State $(chal, reqs) \gets GI$
    \State $(\gamma.state.spk, \gamma.state.ssk) \gets \mathsf{S.GenKeyPair}(\gamma.sec)$
    \State $header \gets (chal, \gamma.state.spk)$
    \State $P \gets \mathsf{ABC.Prove}(\gamma.ipk, \gamma.cred, reqs, header)$
    \State $JP_{TBS} \gets \mathsf{CGKA.GenJP}(\gamma.state, jp\mhyphen type)$
    \State $\sigma \gets \mathsf{S.Sign}(\gamma.state.ssk, JP)$
    \State $JP \gets (JP_{TBS}, \sigma)$
    \State $PP \gets (P, JP)$
    \State \Return{$PP$}
\EndFunction

\\

%

\end{algorithmic}
\end{multicols}

\caption{AA-CGKA construction.}
\label{alg:def}
\end{figure*}

\begin{figure*}
\begin{multicols}{2}
\tiny
\begin{algorithmic}

\Function{createAddProposal}{$\gamma, id, PP$}
    \State $(P, (JP_{TBS}, _)) \gets PP$
    \Assert{ $validatePP(\gamma, PP)$}
    \If{$JP_{TBS}.type = \mathsf{external}$}
        \Assert{ $id = \gamma.self$}
    \EndIf
    \State $prop \gets CGKA.Propose(\gamma, \gamma.self, JP_{TBS}.type, JP_{TBS})$
    \State $prop.cred \gets P$
    \State \Return{$prop$}
\EndFunction

\\

\Function{organiseProps}{$P_L$}
    \State $basic\mhyphen props$
    \For{$(prop, P)$ in $P_L$}
        \If{$prop$ is a basic CGKA proposal}
            \State $basic\mhyphen props \gets basic\mhyphen props + prop$
        \ElsIf{$prop$ is a Requirement proposal}
            \State $req\mhyphen props \gets req\mhyphen props + prop$
        \EndIf
    \EndFor
    \State \Return{$basic\mhyphen props, req\mhyphen props$}
\EndFunction

\\

\Function{validateProps}{$\gamma, basic\mhyphen props, req\mhyphen props$}
    \For{$bp$ in $basic\mhyphen props$}
        \If{$bp.type = \mathsf{add} \lor bp.type = \mathsf{external}$}
            \Assert{ $validatePP(\gamma, bp.cred)$}
        \EndIf
    \EndFor
    \For{$rp$ in $req\mhyphen props$}
        \State $(prop_{TBS}, \sigma) \gets prop$
        \State $pk \gets \gamma.state.members[prop_{TBS}.id].spk$
        \Assert{ $\mathsf{S.Verify}(pk, prop_{TBS}, \sigma)$}
    \EndFor
    \State \Return{$true$}
\EndFunction

\columnbreak

\Function{updateGroupChal}{$chal, C$}
    \State $sig \gets C.sig$
    \State $new\mhyphen chal\gets H(chal || sig)$
    \State \Return{$new\mhyphen chal$}
\EndFunction

\\

\Function{updateReqs}{$reqs, req\mhyphen props$}
    \For{$prop$ in $req\mhyphen props$}
        \State $(prop_{TBS}, \_) \gets prop$
        \If{$prop_{TBS}.type = \mathsf{add}$}
            \State $(ID, type, ID_{req}, new\mhyphen req) \gets prop_{TBS}$
            \Assert{ $reqs[ID_{req}]$ does not exist}
            \State $reqs[ID_{req}] \gets new\mhyphen req$
        \ElsIf{$prop_{TBS}.type = \mathsf{remove}$}
            \State $(ID, type, ID_{req}) \gets prop_{TBS}$
            \Delete{ $reqs[ID_{req}]$}
        \ElsIf{$prop_{TBS}.type = \mathsf{update}$}
            \State $(\_, type, ID_{req}, new\mhyphen req) \gets prop_{TBS}$
            \Assert{ $reqs[ID_{req}]$ exists}
            \State $reqs[ID_{req}] \gets new\mhyphen req$
        \EndIf
    \EndFor
    \State \Return{$new\mhyphen reqs$}
\EndFunction

\\

\Function{validatePP}{$\gamma, PP$}
    \State $(P, (JP_{TBS}, \sigma)) \gets PP$
    \State $ipk \gets getIssuerPK(P)$
    \Assert{ $\mathsf{ABC.VerifyProof}(ipk, P, \gamma.reqs, P.header)$}
    \State $(chal, spk) \gets P.header$
    \Assert{ $chal = \gamma.chal$}
    \Assert{ $\mathsf{S.Verify}(pk, JP, JP_{TBS}, \sigma)$}
\EndFunction

\end{algorithmic}
\end{multicols}

\caption{Auxiliary functions for the AA-CGKA construction.}
\label{alg:def-aux}
\end{figure*}

\begin{figure*}
\begin{multicols}{2}
\tiny
\begin{algorithmic}

\Function{init}{$1^\lambda, issuers$}
    \State $I \gets []$
    \For{$issuer$ in $issuers$}
        \State $I[issuer] \gets \mathsf{ABC.KeyGen}(1^\lambda)$
    \EndFor
\EndFunction

\columnbreak

\Function{issueCredential}{$attrs, issuer$}
    \State $(ipk, isk) \gets I[issuer]$
    \State $cred \gets \mathsf{ABC.Issue}(isk, attrs)$
    \State \Return{$(ipk, cred)$}
\EndFunction

\\

\Function{getIssuerPk}{$P$}
    \State $(ipk, isk) \gets I[P.issuer]$
    \State \Return{$ipk$}
\EndFunction

\end{algorithmic}
\end{multicols}
\caption{Abstract PKI functionality.}
\label{alg:def-pki}
\end{figure*}

\paragraph*{\textbf{Initialisation}}

The $\mathsf{Init}$ function initialises the AA-CGKA user state $\gamma$, including the base CGKA state $\gamma.state$. For simplicity, we will assume that the user's attributes are all certified during initialisation. We remark that the $ID$ parameter for the $\mathsf{Init}$ function is not linked to the user's credential and is only required to identify the user inside the AA-CGKA group.

An AA-CGKA group is initialised through the $\mathsf{Create}$ function. At the start, the only member of the group is its creator, which also initialises a signature key pair. The set of requirements for the group is also defined in this stage.

\paragraph*{\textbf{Proposals and Commits}}

Both \textit{base} and \textit{requirement} proposals must be committed through the operation $\mathsf{Commit}$; in this process, they are signed by a member of the CGKA group. The generated commit message $C$ contains the identity of the committer $id$, base CGKA commit $C_{basic}$, the committed proposals that modify requirements $C_{reqs}$, a \textit{Welcome} message $W$ and a signature $\sigma$. 

To apply the changes included in a commit, the group members must execute the $\mathsf{Process}$ operation. For new members, this also means processing the Welcome message and obtaining the set of requirements from the group. First, members of the group must validate that the signatures included in the commit message are valid. After processing the base CGKA commit, the \textit{requirement proposals} are applied, modifying the current set of requirements.




\subsection{Comparison with CGKA}

We now summarise the main differences and contributions that our AA-CGKA scheme introduces over CGKA. 

First, we employ a different definition of users. Whereas in CGKA users are defined by an abstract \textit{identity}, AA-CGKA users are characterised by a set of attributes, which is reflected in the credential they employ to authenticate to the group. Second, the group contains an additional data structure in AA-CGKA: the group's set of requirements. This also introduces an additional layer of proposals to modify the state of the group in order to allow for insertions, deletions and updates to the requirements. Our AA-CGKA scheme introduces a new phase to the $\mathsf{Process}$ operation related to Commit validation: whenever a new user is inserted into the group, existing members validate the user's credential and check that their attributes match the group's requirement. 

Finally, AA-CGKA users do not employ long-term signature keys for their participation in groups. Instead, they generate fresh key pairs for every group they join. This change is required for the Unlinkability security property, described below in more detail. 

We stress that the following points apply even for existing CGKA schemes that employ attribute-based credentials, such as the one conceptualised in \cite{mls_vc_draft}, as they do not ensure compliance with a set of requirements. Furthermore, they require the ABC to be linked to a long-term signature key, which conflicts with the Unlinkability property.

\section{Security} 
\label{sec:security}

In this Section, we will analyse the security of our AA-CGKA protocol. We start by informally formulating the following security requirements that the protocol must meet:

\begin{enumerate}
    \item \textbf{Requirement Integrity}: Only current members of an AA-CGKA group should be able to modify its requirements. Any attempt to add, update or remove requirements by users who are not part of the group should be rejected.
    \item \textbf{Unforgeability}: It should not be possible to successfully join without possessing a credential that satisfies the AA-CGKA group's requirements. 
    \item \textbf{Unlinkability}: It should be impossible to obtain any private information about users (namely, their credentials) from their interactions in an AA-CGKA group.  
\end{enumerate}

Our Requirement Integrity and Unforgeability properties ensure that unauthorised users cannot modify the state of the group, whether it be by modifying its requirements (which would allow them to potentially modify some of the requirements they do not meet in order to enter the group) or by becoming a member without possessing the required credentials. Lastly, our Unlinkability security property is an adaptation of the ABC scheme's property of the same name to the context of an AA-CGKA group, in order to ensure that the solicitors' privacy is maintained. 

We will define our security properties through an attack game between a Challenger and an Adversary. The Challenger will first create the CGKA group, but we will allow the adversary to execute any number of queries to legitimately modify the state of the group. Both the initialisation process and the available queries for the adversary are shown in Figure \ref{alg:sec-q}. 

Every security game starts with a Setup, in which a number of users and their credentials are initialised. The adversary can request the Challenger to execute any individual AA-CGKA algorithm through its respective query. The exception is $\mathsf{Present}$, which would be redundant as a Presentation Package can be obtained through the $Q_{Propose}$ query. The Challenger keeps track of all generated Proposals and Commits; they are respectively referenced through their indexes in the $Q_{Commit}$ and $Q_{Process}$ queries.

\begin{figure*}
\begin{multicols}{2}
\tiny
\begin{algorithmic}

\State \textbf{Setup}$(1^\lambda, users)$
\Statex \makebox[0pt][l]{\rule{0.8\dimexpr\linewidth-\algorithmicindent}{0.4pt}}

    \State $state, epoch \gets \{\}$
    \State $props, comms \gets []$
   \For{$user$ in $users$}
        \State $(id, attrs, issuer) \gets user$
        \State $state[id] \gets \mathsf{Init}(1^\lambda, id, attrs, issuer)$
        \State $epoch[id] = 0$
    \EndFor

\\

\State \boldsymbol{$Q_{Init}$}$(id, attrs, issuer)$
\Statex \makebox[0pt][l]{\rule{0.8\dimexpr\linewidth-\algorithmicindent}{0.4pt}} 
    \Assert{ $state[id]$ does not exist}
    \State $state[id] \gets \mathsf{Init}(1^\lambda, id, attrs, issuer)$
    \State $epoch[id] = 0$

\\

\State \boldsymbol{$Q_{Create}$}$(id, reqs)$
\Statex \makebox[0pt][l]{\rule{0.8\dimexpr\linewidth-\algorithmicindent}{0.4pt}} 

    \State $state[id] \gets \mathsf{Create}(state[id], reqs)$

\\

\State \boldsymbol{$Q_{Propose}$}$(id, id', prop\mhyphen type)$
\Statex \makebox[0pt][l]{\rule{0.8\dimexpr\linewidth-\algorithmicindent}{0.4pt}} 

    \State $PP \gets ()$
    \If{$prop\mhyphen type = \mathsf{add} \lor prop\mhyphen type = \mathsf{external}$}
        \State $GI \gets \mathsf{PublishInfo}(state[id])$
        \State $PP \gets \mathsf{Present}(state[id'], GI, prop\mhyphen type)$
        \State $PP_L \gets PP_L + PP$
    \EndIf
    \State $(state[id], prop) \gets \mathsf{Propose}(state[id], id', prop\mhyphen type, PP)$
    \State $props \gets props + prop$
    \State \Return{$(prop, PP)$}

\columnbreak

\State \boldsymbol{$Q_{ProposeReqs}$}$(id, prop\mhyphen type, i, new\mhyphen req)$
\Statex \makebox[0pt][l]{\rule{0.8\dimexpr\linewidth-\algorithmicindent}{0.4pt}} 

    \State $(state[id], prop) \gets \mathsf{ProposeReqs}(state[id], prop\mhyphen type, i, new\mhyphen req)$
    \State $props \gets props + (epoch[id], prop)$
    \State \Return{$prop$}

\\

\State \boldsymbol{$Q_{Commit}$}$(id, (i_0, ..., i_n))$
\Statex \makebox[0pt][l]{\rule{0.8\dimexpr\linewidth-\algorithmicindent}{0.4pt}} 

    \State $P_L \gets (props[i_0], ... props[i_n])$
    \State $(state[id], C) \gets \mathsf{Commit}(state[id], P_L)$
    \State $comms \gets comms + (epoch[id], C)$
    \State \Return{$C$}

\\

\State \boldsymbol{$Q_{Process}$}$(id, i_{C})$
\Statex \makebox[0pt][l]{\rule{0.8\dimexpr\linewidth-\algorithmicindent}{0.4pt}} 

    \State $(ep, C) \gets comms[i_{C}]$
    \State $(\gamma, ok) \gets \mathsf{Process}(state[id], C)$
    \If{$\lnot ok$} \State \Return{} \EndIf
    \State $state[id] \gets \gamma$
    \State $epoch[id] \gets ep + 1$

\\

\State \boldsymbol{$Q_{PublishInfo}$}$(id)$
\Statex \makebox[0pt][l]{\rule{0.8\dimexpr\linewidth-\algorithmicindent}{0.4pt}} 

    \State $GI \gets \mathsf{PublishInfo}(state[id])$
    \State \Return{$GI$}

\\

%
%

\State \boldsymbol{$Q_{ExposeCred}$}$(id)$
\Statex \makebox[0pt][l]{\rule{0.8\dimexpr\linewidth-\algorithmicindent}{0.4pt}}

    \State $revealed \gets state[id].cred$ 
    \State $Q_{creds} \gets Q_{creds} + revealed$
    \State \Return{$revealed$}

\end{algorithmic}
\end{multicols}
\caption{Available queries for $A$ in the security games of the AA-CGKA scheme.}
\label{alg:sec-q}
\end{figure*}

%
%
%
%

\begin{figure}
\tiny
\begin{algorithmic}

\State \boldsymbol{$RI^A$}$(1^\lambda, users)$
\Statex \makebox[0pt][l]{\rule{0.5\dimexpr\linewidth-\algorithmicindent}{0.4pt}}

    \State $\mathsf{Setup}(1^\lambda, users)$
    \State $(id, C) \gets A^Q(1^\lambda)$

    \If {$\exists (n, C') \in comms$ s.t $n = state[id] -1$}
        \Assert{ $ C \neq C'$}
    \EndIf

    \State $reqs \gets state[id].reqs$
    \State $(\gamma, ok) \gets \mathsf{Process}(state[id], C)$
    \If{$ok \land state[id].reqs \neq reqs$}
        \State $A$ wins the game
    \EndIf

\\

\State \boldsymbol{$Unf^A$}$(1^\lambda, users)$
\Statex \makebox[0pt][l]{\rule{0.5\dimexpr\linewidth-\algorithmicindent}{0.4pt}}

    \State $\mathsf{Setup}(1^\lambda, users)$
    \State $(id, id', PP_A, prop\mhyphen type) \gets A^Q(1^\lambda)$
    \State $prop \gets \mathsf{Propose}(state[id], prop\mhyphen type, id', PP_A)$
    \State $C \gets \mathsf{Commit}(state[id], [prop])$
    \State $k' \gets A^Q(C)$
    \State $members \gets state[id].state.members$
    \State $(state[id], ok) \gets \mathsf{Process}(state[id], C)$
    \If{$state[id].state.members \not\subseteq members$}
        \If{$\not\exists cred \in Q_{creds}$ s.t. $reqsMet(cred.attrs, reqs)$}
            \If{$state[id].state.k = k'$}
                \State $A$ wins the game
            \EndIf
        \EndIf
    \EndIf

\\

\State \boldsymbol{$Unlink^A$}$(1^\lambda, users)$
\Statex \makebox[0pt][l]{\rule{0.5\dimexpr\linewidth-\algorithmicindent}{0.4pt}}

    \State $\mathsf{Setup}(1^\lambda, users)$
    \State $(id_0, id_1, GI) \gets A^Q(1^\lambda)$
    \State $reqs \gets GI.reqs$
    \Assert{ $reqsMet(reqs, state[id_0].attrs) \land reqsMet(reqs, state[id_1].attrs)$}
    \State $b \overset{{\scriptscriptstyle \operatorname{R}}}{\gets} \{0,1\}$
    \State $PP_b \gets \mathsf{Present}(state[id_b], GI)$
    \State $b' \gets A(PP_b)$
    \If {$b' = b$}
        \State $A$ wins the game
    \EndIf

\\

\end{algorithmic}
\caption{Requirement Integrity, Unforgeability and Unlinkability Security games.}
\label{alg:sec-games}
\end{figure}

We formally define each of the aforementioned security properties in Figure \ref{alg:sec-games}. After executing any number of queries, the adversary outputs some information that depends on the security game that is being executed. In the Requirement Integrity and Unforgeability games, the adversary outputs a commit $C$ and the Challenger verifies if the group's requirements have been modified, or if any new member has been inserted into the group, respectively. If so, the adversary wins the game. While these two games model the possibility of injecting an illegitimate message into the group, in the Unlinkability game we represent the ability of identifying a specific user from a Presentation Package generated by them.

\subsection{Requirement Integrity}

The formal definition of the Requirement Integrity security game $RI^A$ is shown in Figure~\ref{alg:sec-games}. After executing any number of queries, the adversary outputs a custom commit to be processed by an user of her choice; if after the call to $\mathsf{Process}$ the AA-CGKA group's requirements have been modified, then the adversary wins the game since she was able to successfully forge a commit message.

In our security proof we allow injections of a commit from a previous epoch in order to model \textit{replay attacks}, in which an attacker tries to inject a commit that was successfully processed in a previous epoch in order to re-apply some change (i.e., removing a requirement) to the AA-CGKA group. We explicitly rule out the situation of the adversary submitting a commit from the current epoch --- that can be legitimately obtained through the $Q_{Commit}$ query ---, since it would not be considered an \textit{injection}. This is checked by the first assertion of the security game. We will prove that the adversary cannot successfully inject a commit modifying the set of requirements, regardless if it is crafted by her of if it is a legitimate commit from a previous epoch.

We define the advantage in breaking Requirement Integrity as 
\begin{equation}
  \label{eq:adv-ri}
    \mathsf{Adv}^{RI}_A(\lambda) = \Pr [\text{$A$ wins}].
\end{equation}
\begin{definition}
An AA-CGKA scheme provides Requirement Integrity if for all efficient adversaries $A$, $\mathsf{Adv}^{RI}_A(\lambda)$ is negligible.
\end{definition}

\begin{theorem}
\label{th:ri}
If the hash function $H$ is collision-resistant and the signature scheme $S$ is EUF-CMA secure, the proposed AA-CGKA scheme provides Requirement Integrity.
\end{theorem}

We include the full security proof in \ref{app:pr_ri}.

\subsection{Unforgeability}

The formal definition of the Unforgeability security game $Unf^A$ is shown in Figure~\ref{alg:sec-games}. We model the two different methods by which the adversary could try to join a group with a forged Presentation Package: through $\mathsf{add}$ proposals or External Joins. After executing any number of queries, the adversary outputs a custom commit to be processed by an user $id$ of her choice. We first check if after the call to $\mathsf{Process}$ the AA-CGKA group's contains more members than before. If true, we also require that the adversary knows the secret of $id$'s current epoch. This last condition also protects against \textit{replay attacks}, in which a legitimate Presentation Package from another user is reused to add them to the group, but without knowing the secrets employed to generate it. We remark that in order to win, we explicitly forbid the adversary from obtaining any credential that could meet the group's requirements. Thus, we assume that in a real-world deployment, these credentials would be kept in secret \cite{failure-1, failure-2} --- revocation is discussed in Section \ref{sec:practical}, which would correct the leakage of a credential.

For our proof, we will consider the list of legitimately generated Presentation Packages $PP_L$. We will first prove that, in order to successfully join the AA-CGKA group, the adversary must necessarily use one of the elements inside the list (i.e. $PP_A \in PP_L$), since any other Presentation Package will be rejected. Then, we will prove that if $PP_A \in PP_L$, then the adversary would not be able to obtain the group secret $k$.

We define the advantage in breaking Unforgeability as 
\begin{equation}
  \label{eq:adv-unf}
    \mathsf{Adv}^{Unf}_A(\lambda) = \Pr [\text{$A$ wins}].
\end{equation}
\begin{definition}

An AA-CGKA scheme provides Unforgeability if for any efficient adversary $A$, $\mathsf{Adv}^{Unf}_A(\lambda)$ is negligible.
\end{definition}

\begin{theorem}
\label{th:unf}
If the CGKA scheme provides Key Indistinguishability, the signature scheme is EUF-CMA secure and the ABC scheme provides Unforgeability, the proposed AA-CGKA scheme provides Unforgeability.
\end{theorem}

We include the full security proof in \ref{app:pr_unf}.

\subsection{Unlinkability}
The formal definition of the Unlinkability security game $Unlink^A$ is shown in Figure~\ref{alg:sec-games}. After executing any number of queries, the adversary outputs a tuple $(id_0, id_1, GI)$ to the Challenger $C$, who generates a Presentation Package $PP_b$ through the function $\mathsf{Present}$ for the credential of either $id_0$ or $id_1$. The adversary wins the game if he correctly guesses which credential was used.

We define the advantage in breaking Unlinkability as 
\begin{equation}
  \label{eq:adv-unlink}
    \mathsf{Adv}^{Unlink}_A(\lambda) = \Pr [\text{$A$ wins}].
\end{equation}
\begin{definition}
An AA-CGKA scheme provides Unlinkability if for any efficient adversary $A$, $\mathsf{Adv}^{Unlink}_A(\lambda)$ is negligible.
\end{definition}

\begin{theorem}
\label{th:unlink}
If the hash function $H$ is a one-way function and the ABC scheme $ABC$ provides Unlinkability, the proposed AA-CGKA scheme provides Unlinkability.
\end{theorem}

We include the full security proof in \ref{app:pr_unlink}.

\section{Implementation}
\label{sec:impl}

In this Section we introduce the implementation of our AA-CGKA scheme into the Messaging Layer Security (MLS) \cite{mls} specification\footnote{Available at \url{https://github.com/SDABIS/aa-cgka}}. Since MLS is the most popular instantiation of a CGKA protocol, we consider that expressing the functionality of our scheme in terms of MLS operations and messages is relevant for its applicability in the real world. MLS employs TreeKEM as its underlying CGKA protocol \cite{treekem}, which has been proven to fulfil the Key Indistinguishability security property \cite{cgka_analysis}.There are multiple implementations of MLS in different platforms. We develop our implementation from the testbed for experimental analysis of MLS defined in \cite{mls_experimental}, which in turn is based on the OpenMLS Rust implementation \cite{openmls}.

\subsection{MLS}

Our AA-CGKA scheme makes use of some mechanisms of MLS to achieve its functionality, such as \textit{GroupInfo} messages and External Joins. We remark that MLS executes External Joins in a slightly different way to our definition: instead of using a Join Package for both $\mathsf{add}$ and $\mathsf{external}$ proposals, in MLS those are separate procedures that employ different data structures. Nevertheless, our security requirements for Join Packages hold for MLS, as they contain a public key whose private counterpart is only known to the user who created it. 

\paragraph*{\textbf{Using MLS extensions}}

The MLS specification defines \textit{Extensions} for adding new functionality to the protocol. The use of these extensions can be negotiated during the development of an MLS group, making our implementation compatible with traditional MLS. We require the following additions to MLS, to be implemented through Extensions: First, the group's set of requirements. This extension needs no be included in the \textit{GroupContext} structure to be accessible both for current group members and potential joiners, which will obtain it through a \textit{GroupInfo} message. Second, Proposal types for managing requirements, as defined for the $\mathsf{ProposeReqs}$ AA-CGKA algorithm. Third, new credential types for ABC support.

\paragraph*{\textbf{Credential types}}

For the rest of this section, we will consider ABC Presentations as a type of MLS credential (not to be mistaken with an ABC credential). In MLS, credentials are presented to group members inside a \textit{LeafNode} data structure; for $\mathsf{add}$ proposals it is contained in a KeyPackage while for External Joins it appears as the first element in the Update Path of the commit. We include the ABC Presentation in this field, since it is a type of MLS credential. The signing key pair contained in its header will be used to sign the Leaf Node.

We test our implementation with two different ABC schemes:

\begin{enumerate}
    \item JSON Web Tokens with Selective Disclosure (SD\_JWT) \cite{sd_jwt_draft}: In this ABC scheme, the issuer generates a \textit{disclosure} for each of the holder's attribute; to generate a Presentation, the holder must accompany every disclosed attribute with its corresponding disclosure. As every Presentation contains the same disclosure, the SD\_JWT scheme does not provide Unlinkability. For our implementation, we employ the Rust library \textit{sd-jwt-rs} \cite{sd_jwt_impl}. 
    \item BBS Signatures for Verifiable Credentials (BBS\_VC) \cite{vc_bbs}: This scheme applies the BBS signature scheme \cite{bbs} to Verifiable Credentials \cite{vc_20} in order to generate derived credentials that selectively  disclose attributes of the original credential. They contain a proof of knowledge of the original credential's signature. The header that must be provided for the proving operation allows us to include the fresh key pair that will be employed for signing in the AA-CGKA group. Since Presentations are Unlinkable in this ABC scheme, this fresh key pair cannot be associated to any of the holder's long-term key material. We employ the \textit{SSI} Rust library \cite{ssi_impl} for our implementation.
\end{enumerate}

We will compare the performance and behaviour of the  aforementioned ABC schemes to a \textit{trivial solution}, which employs X.509 certificates signed by a trusted Certificate Authority.

The MLS ciphersuite \textit{MLS\_128\_DHKEMX25519\_AES128GCM\_SHA256\_Ed25519} \cite{mls} is employed for the following experiments. We also employ the same algorithms for the other cryptographic primitives of the AA-CGKA scheme as defined in Figure \ref{alg:def}: Ed25519 for the signature scheme $\mathsf{S}$ and SHA256 for the hash function $H$.

\subsection{Performance evaluation}

In this Section, we will evaluate the performance of our implementation and compare it to the trivial solution. We measure the performance of operations that add new members to the group, represented by their \textit{commit type} --- either $\mathsf{add}$ or $\mathsf{external}$. The size of data structures that contain credentials --- KeyPackages, GroupInfo and Welcome messages --- is also studied. We omit analysis of operations such as updates and removals that do not involve the use of credentials, as they remain unaltered in our scheme. We also analyse the impact of group size up to up to 250 group members. Data shown in this section are the result of 10 independent executions for every ABC scheme and commit type.

\begin{figure*}[t]
    \centering
    \begin{subfigure}{0.45\textwidth}
        \centering
        \includegraphics[width=\linewidth]{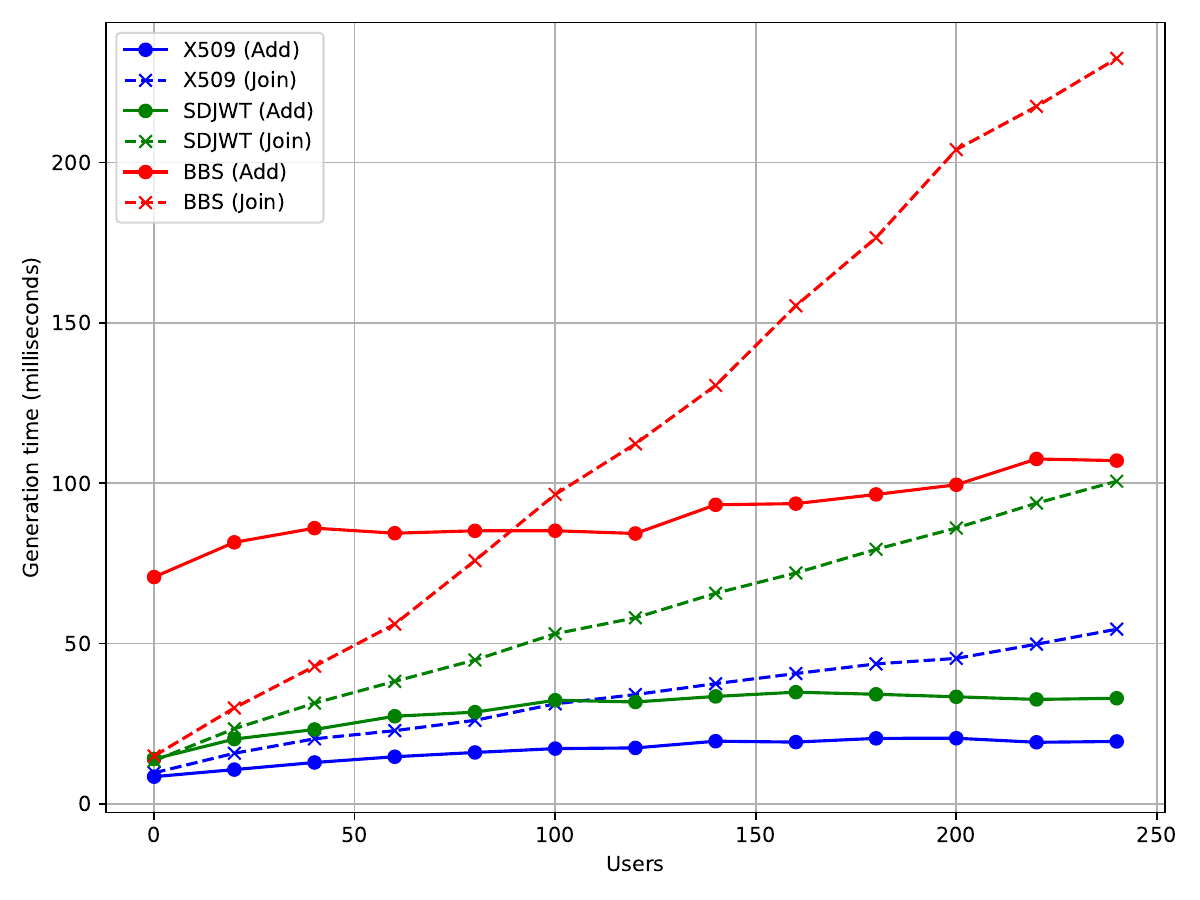}
        \caption{Commit generation time}
        \label{fig:gen}
    \end{subfigure}
    \begin{subfigure}{0.45\textwidth}
        \centering
        \includegraphics[width=\linewidth]{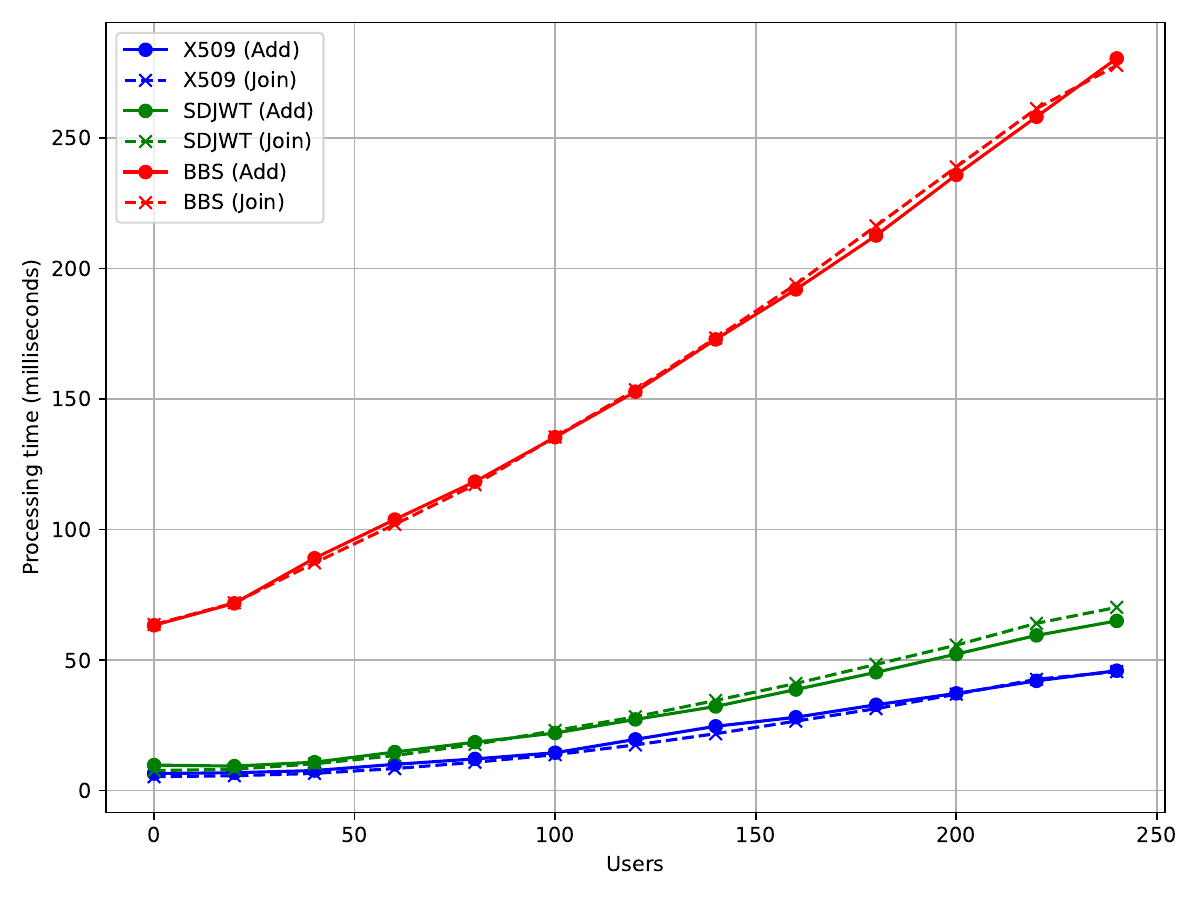}
        \caption{Commit process time}
        \label{fig:proc}
    \end{subfigure}
    \caption{Generation and process time for different ABC schemes as the number of users grow.}
    \label{fig:time}
\end{figure*}

Figure~\ref{fig:gen} shows the time to generate commits for adding new members through $\mathsf{add}$ proposals or External Joins. While SD\_JWT introduces limited overhead, the BBS\_VC scheme has a significantly higher cost --- up to 5 times compared to the trivial approach. While for $\mathsf{add}$ proposals most of the cost is determined by the validation of the added user's credential, for External Joins the most significant time cost is parsing the GroupInfo message --- which contains the full Ratchet Tree of the group.  Commit processing times shown in Figure \ref{fig:proc} are similar to generation times; however, there is not noticeable difference between processing $\mathsf{add}$ and $\mathsf{join}$ proposals.

\begin{table}[t]
    \centering
    \begin{tabular}{lccc}
        \toprule
        & \textbf{Base} & \textbf{SD-JWT} & \textbf{BBS-VC} \\
        \midrule
        \textbf{Gen. Time (ms)}  & 13720  & 14870  & 170320  \\
        \textbf{Size (Bytes)} & 362 & 2880 & 6079 \\
        \bottomrule
    \end{tabular}
    \caption{KeyPackage generation time and size for different ABC schemes.}
    \label{tab:kp}
\end{table}

Table \ref{tab:kp} shows the KeyPackage generation time and size; we omit a comparison with group size as our experimental analysis shows no correlation between the two metrics. As mentioned, it is in this step where Presentation Packages are generated for both commit types. As for other metrics, the cost introduced by BBS\_VC is noticeably higher than the other schemes.

\begin{figure*}[t]
    \centering
    \begin{subfigure}[t]{0.45\textwidth}
        \centering
        \includegraphics[width=\linewidth]{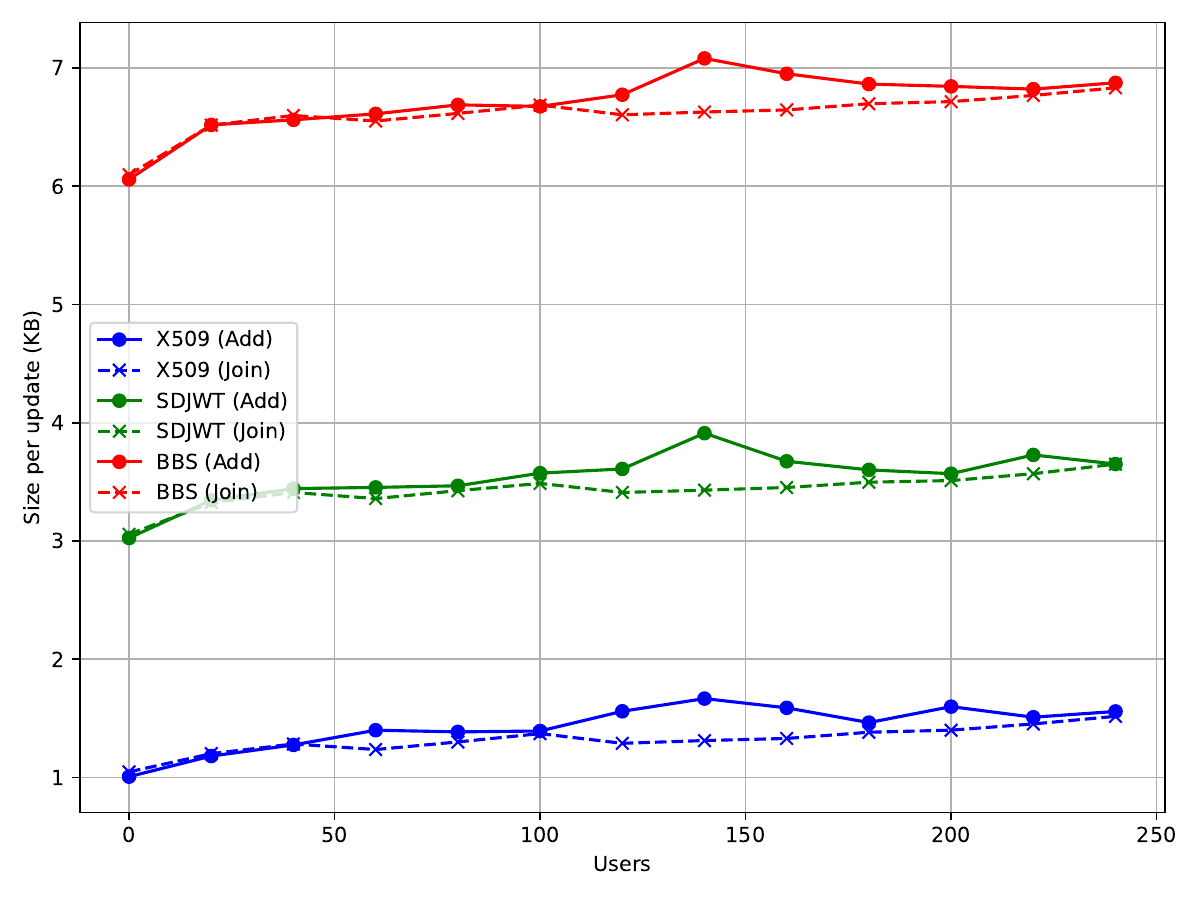}
        \caption{Commit messages.}
        \label{fig:size_commit}
    \end{subfigure}
    \begin{subfigure}[t]{0.45\textwidth}
        \centering
        \includegraphics[width=\linewidth]{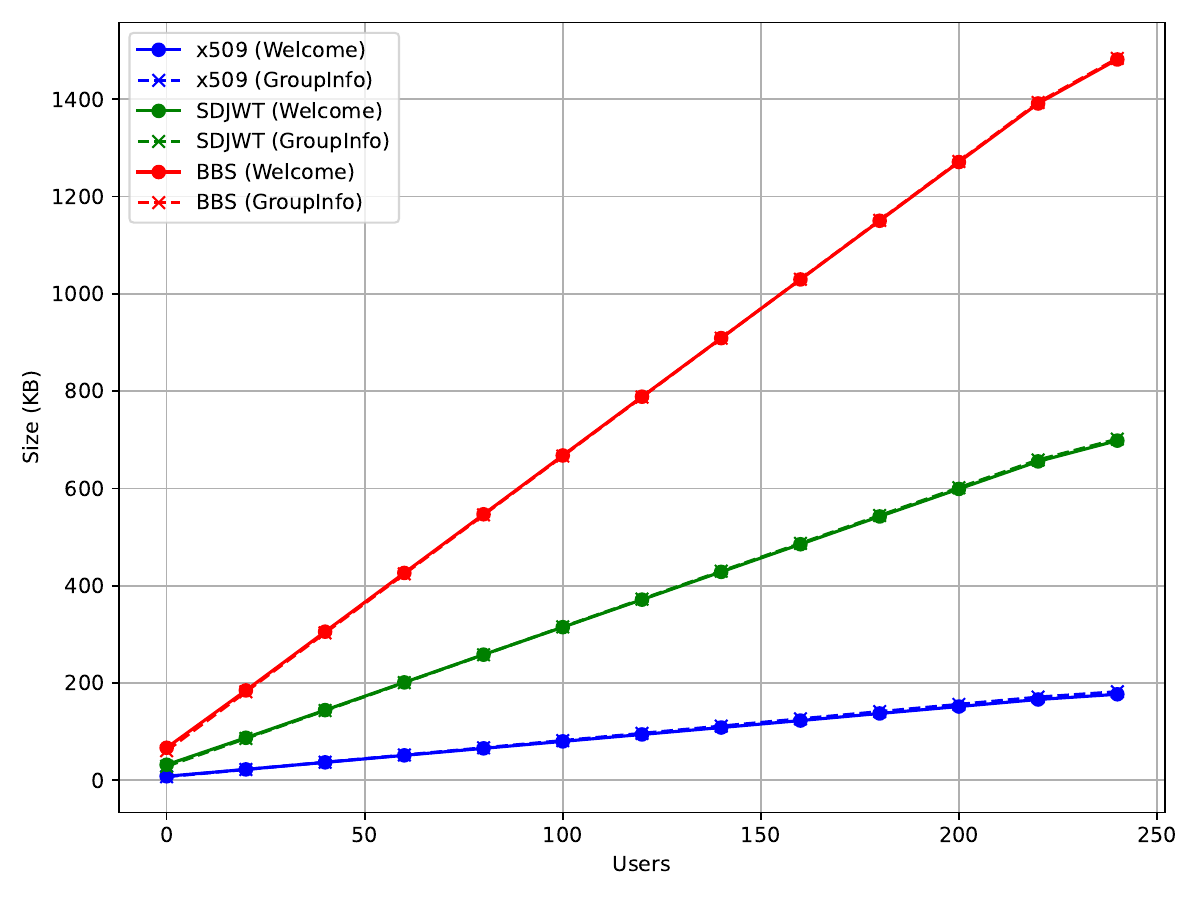}
        \caption{Welcome messages and GroupInfo packages.}
        \label{fig:size_wel}
    \end{subfigure}
    \caption{Message size for different ABC schemes as the number of users grow.}
    \label{fig:size}
\end{figure*}

Figure \ref{fig:size_commit} shows the size of commit messages for the three credential types: it grows slightly as the numbers of users increases, but the credential contained in the $\mathsf{add}$ proposal is the most significant cost. As shown in Figure \ref{fig:size_wel}, both GroupInfo and Welcome message size grows significantly as more Presentations are stored in leaf nodes of the Ratchet Tree. 

\begin{figure}
    \centering
    \includegraphics[width=\linewidth]{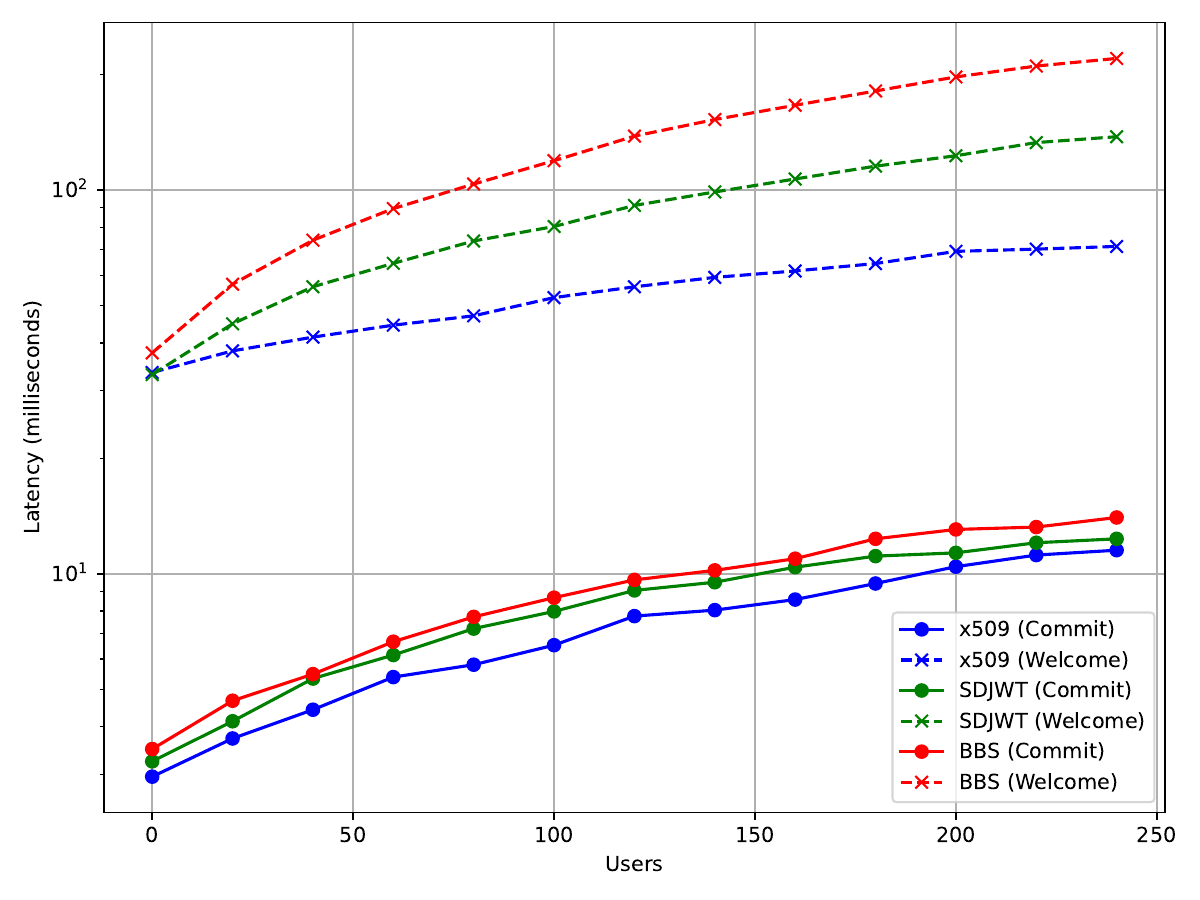}
    \caption{Mean latency for different ABC schemes as the number of users grow.}
    \label{fig:lat}
\end{figure}

We analyse the impact of message size by measuring the \textit{latency}, defined as \textit{the mean time between the instant a Commit is sent and the instant it is processed by its recipient}. Figure \ref{fig:lat} shows the latency for both Commit and Welcome messages; the far bigger size of the latter results in a latency an order of magnitude above the former.

\section{Discussion}
\label{sec:discussion}

In this Section we analyse the results presented in the rest of this work and how they can be applied to real-world environments.

\subsection{Experimental Results}

\begin{table}
    \begin{tabular*}{\textwidth}{@{\extracolsep\fill}c|cc}
    \toprule
    
                          & JWTs with Selective Disclosure (SD\_JWT)     & VCs with BBS   (BBS\_VC)    \\ 
    \midrule
    Presentation          & Reveal disclosures & PoK of signature  \\ 
    Selective Disclosure  & Yes                                       & Yes                                               \\ 
    Unforgeability        & Yes (long term key material) & Yes \\
    Unlinkability         & No & Yes \\
    Computational Cost    & Low                                      & High \\
    Presentation Size     & $\approx$ 2,5 KB                         & $\approx$ 6 KB
    \\ 
    \bottomrule
    \end{tabular*}
\caption{Comparison between the tested ABC schemes. Presentation size is calculated for our tests as it is dependent on the size of the original credential and the amount of disclosed attributes.}
\label{table:comparison}
\end{table}

Table \ref{table:comparison} shows a high-level comparison between the chosen ABC schemes. While all are Unforgeable (SD\_JWT Presentations reference the holder's long term key material), only BBS fulfils the Unlinkability security property ---at the cost of significantly reduced efficiency. While SD\_JWT does not meet said property, it could be useful if high performance ---particularly low processing time--- is required. Nevertheless, our measurements show computing costs of less that half a second even for large groups of 250 members.

One of the most significant costs introduced in our scheme is related to message size; as shown in Figure \ref{fig:size_wel}, GroupInfo and Welcome messages that contain the full Ratchet Tree reach sizes of 1MB for 250 group members. This is specially relevant for BBS\_VC Presentations as they contain lengthy proofs of knowledge. Indeed, the size of these messages also impacts computational costs such as External Join generation time. We stress that these messages only need to be processed by the joiner and thus would not impact the bandwidth of most group members. 

The size of Welcome messages has a noticeable impact on latency. The time to distribute them is approximately 10 times larger than the latency for regular Commits. We note that, although using ABCs as credentials --- particularly BBS\_VC --- increases message size, this cost is mostly unavoidable as the trivial solution is in the same order of magnitude.

\subsection{Practical Considerations}
\label{sec:practical}
We now discuss aspects of our protocol that are relevant to a real-world deployment, particularly for distributed environments:

\begin{itemize}
    \item Bandwidth-limited devices: Although Welcome messages are one-time and only the joiner needs to process them, their 1MB size may be too high for bandwidth-constrained devices. There exist solutions that reduce the size of the Welcome message by not including the full Ratchet Tree \cite{mls_partial_draft} which dramatically reduces their size, but prevents members from issuing Commits. This tradeoff may be acceptable for users only interested in receiving updates.
    \item Issuer discoverability: In a distributed settings, users may not recognise the same set of issuers as valid. A group may communicate a list of acceptable issuers through its set of requirements or by otherwise including them in the GroupInfo package.
    \item Credential Revocation: In real-world environments, verifiers also need to check that the received credential is not revoked, which is complicated by the Unlinkability property of the ABC scheme. Since the verifier never gets access to the original credential, the holder must then prove that the credential is not revoked. This can be solved through an additional proof-of-membership \cite{zkcreds} to an issuer-controlled accumulator, that can be added to the Presentation's header or otherwise inserted into MLS through a KeyPackage Extension. Users may also be required to periodically commit an Update proposal to prove their credential is still valid. 
    \item Further Commit validation: CGKA is inherently vulnerable to insider attacks that fragment the group into incompatible states \cite{itk}; in order to prevent them, an additional consensus layer could be introduced that ensures that all parties agree on the next commit \cite{discreet}, using lightweight protocols such as \textit{Context Adaptive Cooperation} \cite{cac}. There also exist CGKA variants --- that are compatible with our AA-CGKA construction --- that restrict which users are able to issue commits \cite{a_cgka, cgka_fa}.
\end{itemize}

\subsection{Real-world Applications}

Our proposed AA-CGKA scheme is useful for the establishment of online communities between users that wish to retain their privacy. Groups can specify a list of attributes that all members must possess, such that communities can be easily formed between users with common characteristics or interests. Furthermore, users only need to prove the specific attributes required by the group and their actions across multiple groups are unlinkable, ensuring that they can safely interact online even in untrusted environments.

Among other applications, AA-CGKA schemes can be useful in healthcare environments. This would allow for the creation of AA-CGKA groups between patients and medical experts, each with their own credentials proving their role in the group. The security properties provided by the AA-CGKA scheme ensure that participants can safely discuss sensitive topics between each other and with experts, which may serve as a preliminary step towards receiving more direct assistance.

The establishment of attribute-gated online communities is also useful for the protection of their members. Indeed, unregulated anonymous networks may present dangers for its participants: in the case of online communities populated by minors, they can be exposed to adult groomers \cite{grooming-2} concealing their age \cite{grooming-1}. Using an AA-CGKA scheme to enforce an age range could prevent these intrusions.

\section{Conclusion and Future Work}
\label{sec:conclusion}

In this work, we have presented the Attribute-Authenticated Continuous Group Key Agreement (AA-CGKA) scheme. In this protocol, members of a group can specify a set of attribute requirements that all users aiming to join the group must fulfil. To authenticate themselves solicitors must present an ABC proving they meet the imposed requirements. The use of Selective Disclosure and credential derivation allows users to only reveal the minimum private information required while maintaining the trustworthiness of their credentials, and prevents eavesdroppers from linking their activities through different AA-CGKA groups. We have introduced the security properties of Requirement Integrity, Unforgeability and Unlinkability for our AA-CGKA protocol and provided formal security proofs.

Our scheme allows for the creation of messaging groups between users who possess a common attribute but do not desire to reveal any other information about themselves. This is specially suited for establishing communication in decentralised environments, in which privacy is an important concern. Furthermore, the use of \textit{External Joins} allows users to join a group without the necessity of invitation. Thus, we envision AA-CGKA groups as having a dynamic composition, with users joining and leaving at a high pace.

\section*{Declarations}

\subsection*{Availability of data and materials}

We publish the implementation of our AA-CGKA scheme as open source:

\begin{itemize}
    \item Name: \textit{aa-cgka}
    \item Home page: \url{https://github.com/SDABIS/aa-cgka}
    \item Archived version: N/A
    \item Operating system: Platform independent
    \item Programming language: Rust
    \item Other requirements: Docker
    \item License: MIT
    \item Any restrictions: N/A
\end{itemize}

\subsection*{Funding}

D.S. acknowledges support from Xunta de Galicia and the European Union (European Social Fund - ESF) scholarship [ED481A-2023-219].

This work is funded by the Plan Complementario de Comunicaciones Cu\'anticas, Spanish Ministry of Science and Innovation(MICINN), Plan de Recuperación NextGeneration, European Union (PRTR-C17.I1, CITIC Ref. 305.2022), and Regional Government of Galicia (Agencia Gallega de Innovación, GAIN, CITIC Ref. 306.2022).

\subsection*{Authors' contributions}

David Soler: Conceptualization, Formal Analysis, Investigation, Software, Writing - original draft
Carlos Dafonte: Conceptualization, Supervision, Writing - review \& editing
Manuel Fernández-Veiga: Conceptualization, Supervision, Writing - review \& editing"
Ana Fernández Vilas: Conceptualization, Supervision, Writing - review \& editing"
Francisco J. Nóvoa: Conceptualization, Supervision, Writing - review \& editing"

\subsection*{Acknowledgements}

Not Applicable.

\bibliographystyle{elsarticle-num} 
\bibliography{main}

@misc{a_cgka,
      author = {David Balbás and Daniel Collins and Serge Vaudenay},
      title = {Cryptographic Administration for Secure Group Messaging},
      howpublished = {Cryptology ePrint Archive, Paper 2022/1411},
      year = {2022},
}

@misc{mls_experimental,
      title={Experimental Analysis of Efficiency of the Messaging Layer Security for Multiple Delivery Services}, 
      author={David Soler and Carlos Dafonte and Manuel Fernández-Veiga and Ana Fernández Vilas and Francisco J. Nóvoa},
      year={2025},
      eprint={2502.18303},
      archivePrefix={arXiv},
      primaryClass={cs.CR},
      url={https://arxiv.org/abs/2502.18303}, 
}

@misc{cgka_analysis_etk,
      author = {Cas Cremers and Esra Günsay and Vera Wesselkamp and Mang Zhao},
      title = {{ETK}: External-Operations {TreeKEM} and the Security of {MLS} in {RFC} 9420},
      howpublished = {Cryptology {ePrint} Archive, Paper 2025/229},
      year = {2025},
      url = {https://eprint.iacr.org/2025/229}
}

@ARTICLE{failure-1,
  author={Wang, Qingxuan and Wang, Ding},
  journal={IEEE Transactions on Information Forensics and Security}, 
  title={Understanding Failures in Security Proofs of Multi-Factor Authentication for Mobile Devices}, 
  year={2023},
  volume={18},
  number={},
  pages={597-612},
  doi={10.1109/TIFS.2022.3227753}}

@article{failure-2,
  author    = {Neal Koblitz and Alfred J. Menezes},
  title     = {Another Look at ``Provable Security''},
  journal   = {Journal of Cryptology},
  year      = {2007},
  volume    = {20},
  number    = {1},
  pages     = {3--37},
  doi       = {10.1007/s00145-005-0432-z},
  url       = {https://doi.org/10.1007/s00145-005-0432-z},
  issn      = {1432-1378},
}

@incollection{cgka_analysis,
	address = {Cham},
	title = {Security {Analysis} and {Improvements} for the {IETF} {MLS} {Standard} for {Group} {Messaging}},
	volume = {12170},
	isbn = {978-3-030-56783-5 978-3-030-56784-2},
	language = {en},
	urldate = {2024-01-31},
	booktitle = {Advances in {Cryptology} – {CRYPTO} 2020},
	publisher = {Springer International Publishing},
	author = {Alwen, Joël and Coretti, Sandro and Dodis, Yevgeniy and Tselekounis, Yiannis},
	editor = {Micciancio, Daniele and Ristenpart, Thomas},
	year = {2020},
	doi = {10.1007/978-3-030-56784-2\_9},
	note = {Series Title: Lecture Notes in Computer Science},
	pages = {248--277},
}

@misc{modular,
      author = {Joël Alwen and Sandro Coretti and Yevgeniy Dodis and Yiannis Tselekounis},
      title = {Modular Design of Secure Group Messaging Protocols and the Security of {MLS}},
      howpublished = {Cryptology {ePrint} Archive, Paper 2021/1083},
      year = {2021},
      url = {https://eprint.iacr.org/2021/1083}
}

@inproceedings{urs,
  title={Composable and modular anonymous credentials: Definitions and practical constructions},
  author={Camenisch, Jan and Dubovitskaya, Maria and Haralambiev, Kristiyan and Kohlweiss, Markulf},
  booktitle={Advances in Cryptology--ASIACRYPT 2015: 21st International Conference on the Theory and Application of Cryptology and Information Security, Auckland, New Zealand, November 29--December 3, 2015, Proceedings, Part II 21},
  pages={262--288},
  year={2015},
  organization={Springer}
}

@article{ssi_survey,
  title={A survey of self-sovereign identity ecosystem},
  author={Soltani, Reza and Nguyen, Uyen Trang and An, Aijun},
  journal={Security and Communication Networks},
  volume={2021},
  pages={1--26},
  year={2021},
  publisher={Hindawi Limited}
}

@inproceedings{abc_fw,
  title={A cryptographic framework for the controlled release of certified data},
  author={Bangerter, Endre and Camenisch, Jan and Lysyanskaya, Anna},
  booktitle={Security Protocols: 12th International Workshop, Cambridge, UK, April 26-28, 2004. Revised Selected Papers 12},
  pages={20--42},
  year={2006},
  organization={Springer}
}

@incollection{cl,
	address = {Berlin, Heidelberg},
	title = {Signature {Schemes} and {Anonymous} {Credentials} from {Bilinear} {Maps}},
	volume = {3152},
	isbn = {978-3-540-22668-0 978-3-540-28628-8},
	language = {en},
	urldate = {2024-02-02},
	booktitle = {Advances in {Cryptology} – {CRYPTO} 2004},
	publisher = {Springer Berlin Heidelberg},
	author = {Camenisch, Jan and Lysyanskaya, Anna},
	editor = {Hutchison, David and Kanade, Takeo and Kittler, Josef and Kleinberg, Jon M. and Mattern, Friedemann and Mitchell, John C. and Naor, Moni and Nierstrasz, Oscar and Pandu Rangan, C. and Steffen, Bernhard and Sudan, Madhu and Terzopoulos, Demetri and Tygar, Dough and Vardi, Moshe Y. and Weikum, Gerhard and Franklin, Matt},
	year = {2004},
	doi = {10.1007/978-3-540-28628-8\_4},
	note = {Series Title: Lecture Notes in Computer Science},
	pages = {56--72},
}

@incollection{itk,
	address = {Cham},
	title = {On the {Insider} {Security} of {MLS}},
	volume = {13508},
	isbn = {978-3-031-15978-7 978-3-031-15979-4},
	language = {en},
	urldate = {2024-01-31},
	booktitle = {Advances in {Cryptology} – {CRYPTO} 2022},
	publisher = {Springer Nature Switzerland},
	author = {Alwen, Joël and Jost, Daniel and Mularczyk, Marta},
	editor = {Dodis, Yevgeniy and Shrimpton, Thomas},
	year = {2022},
	doi = {10.1007/978-3-031-15979-4\_2},
	note = {Series Title: Lecture Notes in Computer Science},
	pages = {34--68},
}

@misc{decaf,
      author = {Joël Alwen and Benedikt Auerbach and Miguel Cueto Noval and Karen Klein and Guillermo Pascual-Perez and Krzysztof Pietrzak},
      title = {DeCAF: Decentralizable Continuous Group Key Agreement with Fast Healing},
      howpublished = {Cryptology ePrint Archive, Paper 2022/559},
      year = {2022},
}

@article{treekem,
	title = {{TreeKEM}: {Asynchronous} {Decentralized} {Key} {Management} for {Large} {Dynamic} {Groups}},
	language = {en},
	author = {Bhargavan, Karthikeyan and Barnes, Richard and Rescorla, Eric},
    year = {2018},
    journal = {.},
}

@inproceedings{saik,
	address = {Los Angeles CA USA},
	title = {Server-{Aided} {Continuous} {Group} {Key} {Agreement}},
	isbn = {978-1-4503-9450-5},
	doi = {10.1145/3548606.3560632},
	language = {en},
	urldate = {2024-01-31},
	booktitle = {Proceedings of the 2022 {ACM} {SIGSAC} {Conference} on {Computer} and {Communications} {Security}},
	publisher = {ACM},
	author = {Alwen, Joël and Hartmann, Dominik and Kiltz, Eike and Mularczyk, Marta},
	month = nov,
	year = {2022},
	pages = {69--82},
}

@inproceedings{dec_ack,
	address = {Virtual Event Republic of Korea},
	title = {Key {Agreement} for {Decentralized} {Secure} {Group} {Messaging} with {Strong} {Security} {Guarantees}},
	isbn = {978-1-4503-8454-4},
	doi = {10.1145/3460120.3484542},
	language = {en},
	urldate = {2024-01-31},
	booktitle = {Proceedings of the 2021 {ACM} {SIGSAC} {Conference} on {Computer} and {Communications} {Security}},
	publisher = {ACM},
	author = {Weidner, Matthew and Kleppmann, Martin and Hugenroth, Daniel and Beresford, Alastair R.},
	month = nov,
	year = {2021},
	pages = {2024--2045},
}

@incollection{fork,
	address = {Cham},
	title = {Fork-{Resilient} {Continuous} {Group} {Key} {Agreement}},
	volume = {14084},
	isbn = {978-3-031-38550-6 978-3-031-38551-3},
	language = {en},
	urldate = {2024-01-31},
	booktitle = {Advances in {Cryptology} – {CRYPTO} 2023},
	publisher = {Springer Nature Switzerland},
	author = {Alwen, Joël and Mularczyk, Marta and Tselekounis, Yiannis},
	editor = {Handschuh, Helena and Lysyanskaya, Anna},
	year = {2023},
	doi = {10.1007/978-3-031-38551-3\_13},
	note = {Series Title: Lecture Notes in Computer Science},
	pages = {396--429},
}

@incollection{bounds,
	address = {Cham},
	title = {On the {Price} of {Concurrency} in {Group} {Ratcheting} {Protocols}},
	volume = {12551},
	isbn = {978-3-030-64377-5 978-3-030-64378-2},
	language = {en},
	urldate = {2024-01-31},
	booktitle = {Theory of {Cryptography}},
	publisher = {Springer International Publishing},
	author = {Bienstock, Alexander and Dodis, Yevgeniy and Rösler, Paul},
	editor = {Pass, Rafael and Pietrzak, Krzysztof},
	year = {2020},
	doi = {10.1007/978-3-030-64378-2\_8},
	note = {Series Title: Lecture Notes in Computer Science},
	pages = {198--228},
}

@inproceedings{cgka_fa,
	address = {Charlotte NC USA},
	title = {Continuous {Group} {Key} {Agreement} with {Flexible} {Authorization} and {Its} {Applications}},
	isbn = {9798400700996},
	doi = {10.1145/3579987.3586570},
	language = {en},
	urldate = {2024-01-31},
	booktitle = {Proceedings of the 9th {ACM} {International} {Workshop} on {Security} and {Privacy} {Analytics}},
	publisher = {ACM},
	author = {Kajita, Kaisei and Emura, Keita and Ogawa, Kazuto and Nojima, Ryo and Ohtake, Go},
	month = apr,
	year = {2023},
	pages = {3--13},
}

@inproceedings{art,
	address = {Toronto Canada},
	title = {On {Ends}-to-{Ends} {Encryption}: {Asynchronous} {Group} {Messaging} with {Strong} {Security} {Guarantees}},
	isbn = {978-1-4503-5693-0},
	shorttitle = {On {Ends}-to-{Ends} {Encryption}},
	doi = {10.1145/3243734.3243747},
	language = {en},
	urldate = {2024-01-31},
	booktitle = {Proceedings of the 2018 {ACM} {SIGSAC} {Conference} on {Computer} and {Communications} {Security}},
	publisher = {ACM},
	author = {Cohn-Gordon, Katriel and Cremers, Cas and Garratt, Luke and Millican, Jon and Milner, Kevin},
	month = oct,
	year = {2018},
	pages = {1802--1819},
}

@incollection{issuer_hiding,
	address = {Cham},
	title = {Issuer-{Hiding} {Attribute}-{Based} {Credentials}},
	volume = {13099},
	isbn = {978-3-030-92547-5 978-3-030-92548-2},
	language = {en},
	urldate = {2024-02-05},
	booktitle = {Cryptology and {Network} {Security}},
	publisher = {Springer International Publishing},
	author = {Bobolz, Jan and Eidens, Fabian and Krenn, Stephan and Ramacher, Sebastian and Samelin, Kai},
	editor = {Conti, Mauro and Stevens, Marc and Krenn, Stephan},
	year = {2021},
	doi = {10.1007/978-3-030-92548-2\_9},
	note = {Series Title: Lecture Notes in Computer Science},
	pages = {158--178},
}

@incollection{cgka_anon,
	address = {Cham},
	title = {Membership {Privacy} for {Asynchronous} {Group} {Messaging}},
	volume = {13720},
	isbn = {978-3-031-25658-5 978-3-031-25659-2},
	language = {en},
	urldate = {2024-01-31},
	booktitle = {Information {Security} {Applications}},
	publisher = {Springer Nature Switzerland},
	author = {Emura, Keita and Kajita, Kaisei and Nojima, Ryo and Ogawa, Kazuto and Ohtake, Go},
	editor = {You, Ilsun and Youn, Taek-Young},
	year = {2023},
	doi = {10.1007/978-3-031-25659-2\_10},
	note = {Series Title: Lecture Notes in Computer Science},
	pages = {131--142},
}

@techreport{mls,
	type = {Request for {Comments}},
	title = {The {Messaging} {Layer} {Security} ({MLS}) {Protocol}},
	number = {RFC 9420},
	urldate = {2024-02-01},
	institution = {Internet Engineering Task Force},
	author = {Barnes, Richard and Beurdouche, Benjamin and Robert, Raphael and Millican, Jon and Omara, Emad and Cohn-Gordon, Katriel},
	month = jul,
	year = {2023},
	doi = {10.17487/RFC9420},
	note = {Num Pages: 132},
}

@inproceedings{zkcreds,
  title={zk-creds: Flexible anonymous credentials from zksnarks and existing identity infrastructure},
  author={Rosenberg, Michael and White, Jacob and Garman, Christina and Miers, Ian},
  booktitle={2023 IEEE Symposium on Security and Privacy (SP)},
  pages={790--808},
  year={2023},
  organization={IEEE}
}

@techreport{mls_vc_draft,
	type = {Internet {Draft}},
	title = {Additional {MLS} {Credentials}},
	number = {draft-barnes-mls-addl-creds-00},
	urldate = {2024-02-01},
	institution = {Internet Engineering Task Force},
	author = {Barnes, Richard and Nandakumar, Suhas},
	month = jul,
	year = {2023},
	note = {Num Pages: 12},
}

@techreport{mls_partial_draft,
	type = {Internet {Draft}},
	title = {Partial MLS},
	number = { draft-ietf-mls-partial-00},
	urldate = {2025-12-16},
	institution = {Internet Engineering Task Force},
	author = {Kiefer, Franziskus and Bhargavan, Karthikeyan and Barnes, Richard and Alwen, Joel and Mularczyk, Marta},
	month = jul,
	year = {2023},
	note = {Num Pages: 12},
}

@misc{cac,
      title={Context Adaptive Cooperation}, 
      author={Timothé Albouy and Davide Frey and Mathieu Gestin and Michel Raynal and François Taïani},
      year={2024},
      eprint={2311.08776},
      archivePrefix={arXiv},
      primaryClass={cs.DC},
      url={https://arxiv.org/abs/2311.08776}, 
}

@article{discreet,
  author    = {Ludovic Paillat and Claudia-Lavinia Ignat and Davide Frey and Mathieu Turuani and Amine Ismail},
  title     = {Discreet: distributed delivery service with context-aware cooperation},
  journal   = {Annals of Telecommunications},
  year      = {2025},
  volume    = {80},
  number    = {3},
  pages     = {357--374},
  doi       = {10.1007/s12243-024-01053-1},
  issn      = {1958-9395}
}

@inproceedings{abs-original,
  title={Attribute-based signatures},
  author={Maji, Hemanta K and Prabhakaran, Manoj and Rosulek, Mike},
  booktitle={Cryptographers’ track at the RSA conference},
  pages={376--392},
  year={2011},
  organization={Springer}
}

@inproceedings{abs-1,
  title={Attribute-based signatures for supporting anonymous certification},
  author={Kaaniche, Nesrine and Laurent, Maryline},
  booktitle={European symposium on research in computer security},
  pages={279--300},
  year={2016},
  organization={Springer}
}

@inproceedings{abs-2,
  title={Attribute-based signatures with user-controlled linkability without random oracles},
  author={El Kaafarani, Ali and Ghadafi, Essam},
  booktitle={IMA International Conference on Cryptography and Coding},
  pages={161--184},
  year={2017},
  organization={Springer}
}

@TechReport{vc_20,
  author      = "Gabe Cohen and Ted Thibodeau and Ivan Herman and Manu Sporny and Michael Jones",
  title       = "Verifiable Credentials Data Model v2.0",
  month       = feb,
  note        = "https://www.w3.org/TR/2024/CRD-vc-data-model-2.0-20240207/",
  year        = "2024",
  type        = "Candidate Recommendation",
  institution = "W3C",
}

@inproceedings{suf-cma,
  title={The provable security of ed25519: theory and practice},
  author={Brendel, Jacqueline and Cremers, Cas and Jackson, Dennis and Zhao, Mang},
  booktitle={2021 IEEE Symposium on Security and Privacy (SP)},
  pages={1659--1676},
  year={2021},
  organization={IEEE}
}

@Misc{openmls,
  author       = {PhoenixR\&D},
  howpublished = {\url{https://github.com/openmls/openmls}},
  title        = {OpenMLS},
  year         = {2024},
  commit       = {1007599},
  journal      = {GitHub repository},
  publisher    = {GitHub},
}

@TechReport{vc_bbs,
  author      = "Greg Bernstein and Manu Sporny",
  title       = "Data Integrity BBS Cryptosuites {v1.0}",
  month       = mar,
  note        = "https://www.w3.org/TR/vc-di-bbs/",
  year        = "2024",
  type        = "{W3C} Recommendation",
  institution = "W3C",
}

@techreport{bbs,
	type = {Internet {Draft}},
	title = {The {BBS} {Signature} {Scheme}},
	url = {https://datatracker.ietf.org/doc/draft-irtf-cfrg-bbs-signatures},
	number = {draft-irtf-cfrg-bbs-signatures-05},
	urldate = {2024-02-02},
	institution = {Internet Engineering Task Force},
	author = {Looker, Tobias and Kalos, Vasilis and Whitehead, Anew and Lodder, Mike},
	month = dec,
	year = {2023},
	note = {Num Pages: 115},
}

@techreport{sd_jwt_draft,
	type = {Internet {Draft}},
	title = {Selective Disclosure for {JWTs (SD-JWT)}},
	url = {https://datatracker.ietf.org/doc/draft-ietf-oauth-selective-disclosure-jwt/},
	number = {draft-ietf-oauth-selective-disclosure-jwt-08},
	urldate = {2024-04-17},
	institution = {Internet Engineering Task Force},
	author = {Fett, Daniel and Yasuda, Kristina and Campbell, Daniel},
	month = jul,
	year = {2023},
	note = {Num Pages: 12},
}

@Misc{sd_jwt_impl,
  author       = {OpenWalletFoundation},
  howpublished = {\url{https://github.com/openwallet-foundation-labs/sd-jwt-rust}},
  title        = {{SD-JWT} Rust Reference Implementation},
  year         = {2024},
  commit       = {308763d},
  journal      = {GitHub repository},
  publisher    = {GitHub},
}

@Misc{ssi_impl,
  author       = {SpruceID},
  howpublished = {\url{https://github.com/spruceid/ssi}},
  title        = {SSI},
  year         = {2024},
  commit       = {976e260},
  journal      = {GitHub repository},
  publisher    = {GitHub},
}

@inproceedings{grooming-1,
  title={A Step Towards Detecting Online Grooming--Identifying Adults Pretending to be Children},
  author={Ashcroft, Michael and Kaati, Lisa and Meyer, Maxime},
  booktitle={2015 European Intelligence and Security Informatics Conference},
  pages={98--104},
  year={2015},
  organization={IEEE}
}

@techreport{grooming-2,
  title        = {Tech Platforms Used by Online Child Sexual Abuse Offenders: Research Report with Actionable Recommendations for the Tech Industry},
  author       = {Insoll, Tegan and Soloveva, Valeriia and Díaz Bethencourt, Eva and Ovaska, Anna and Vaaranen‑Valkonen, Nina},
  year         = {2024},
  institution  = {Protect Children / Suojellaan Lapsia ry},
  type         = {Research Report},
  note         = {Accessed: 25-09-2025}
}

\appendix

\section{Security Games}
\label{app:games}

In this Appendix we include the security games for the cryptographic primitives we employ in our AA-CGKA scheme. They are specially relevant for the works we have extracted these definitions from are mentioned in the relevant sections.

\subsection{Continuous Group Key Agreement}
\label{app:cgka}

The Key Indistinguishability security game for a CGKA scheme is shown in Figure \ref{alg:cgka-kind}. The game is the same as in \cite{cgka_analysis} and \cite{a_cgka}, but with two relevant changes: firstly, the Join Package data structure are made explicit with the function $\mathsf{GenJP}$, which were included in our definition to highlight the authentication process in CGKA groups. Secondly, we include the $\mathsf{external}$ proposal type, which allows for a different type of insertion of members to the CGKA group. External Joins have been recently formalised by \cite{cgka_analysis_etk}, although the authors use a different security models to ours. 
Readers are encouraged to obtain more information in \cite{cgka_analysis, cgka_analysis_etk}, specially that related to the \textit{safe predicate}, which ensures that trivial attacks (such as challenging an epoch in    which the shared secret has already been revealed) are ruled out. 

\subsection{Attribute-Based Credentials}
\label{app:abc}

Figures \ref{alg:abc-unf} and \ref{alg:abc-unlink} show the formal definition of the Unforgeability $Unf^A$ and Unlinkability $Unlink^A$ security games, respectively. These definitions are derived from \cite{issuer_hiding}, but have been adapted to our specific goals. 

As explained in Section \ref{sec:abc}, the Unforgeability security requirement ensures that no attacker can forge a valid presentation without access to a credential. The adversary $A$ is allowed to request the issuance of arbitrary credentials and to generate presentations from them. $A$ cannot gain access to any of the issued credentials unless she specifically asks for them: in that case, $A$ is forbidden from outputting a presentation that could have been generated through that revealed credential.

\begin{figure}
\tiny
\begin{algorithmic}

\State \boldsymbol{$Unf^A$}$(1^\lambda, issuers)$
\Statex \makebox[0pt][l]{\rule{0.5\dimexpr\linewidth-\algorithmicindent}{0.4pt}}

    \State $pubs, privs, issued, presented, revealed \gets []$
    \For{$issuer$ in $issuers$}
        \State $(ipk, isk) \gets \mathsf{ABC.KeyGen}(1^\lambda)$
        \State $pubs \gets pubs + ipk$
        \State $privs \gets privs + isk$
    \EndFor
    \State $(i, P, header) \gets A^Q(pubs)$
    \State $ipk \gets pubs[i]$
    \If{ $P \notin presented$ \\
    $\land \not\exists (i, cred)$ s.t. $disc\mhyphen attrs \in cred.attrs$ \\
    $\land \mathsf{ABC.VerifyProof}(ipk, P, disc\mhyphen attrs, header)$}
        \State $A$ wins the game
    \EndIf

\\

\State \boldsymbol{$Q_{Issue}$}$(i, attrs)$
\Statex \makebox[0pt][l]{\rule{0.5\dimexpr\linewidth-\algorithmicindent}{0.4pt}}

    \State $isk \gets privs[i]$
    \State $cred \gets \mathsf{ABC.Issue}(isk_i, attrs)$
    \State $issued \gets issued + (i, cred)$

\\

\State \boldsymbol{$Q_{Present}$}$(j, disc\mhyphen attrs, header)$
\Statex \makebox[0pt][l]{\rule{0.5\dimexpr\linewidth-\algorithmicindent}{0.4pt}}

    \State $(i, cred) \gets issued[j]$
    \State $ipk \gets pubs[i]$
    \State $P \gets \mathsf{ABC.Prove}(ipk, cred, disc\mhyphen attrs, header)$
    \State $presented \gets presented + P$
    \State \Return{$P$}

\\

\State \boldsymbol{$Q_{Reveal}$}$(j)$
\Statex \makebox[0pt][l]{\rule{0.5\dimexpr\linewidth-\algorithmicindent}{0.4pt}}

    \State $(i, cred) \gets issued[j]$
    \State $revealed \gets revealed + cred$
    \State \Return{$(i, cred)$}

\end{algorithmic}
\caption{Unforgeability Security game for an ABC scheme.}
\label{alg:abc-unf}
\end{figure}

In the Unlinkability game, $A$'s objective is to guess from which credential a specific presentation was generated. This is relevant for privacy considerations, as this property ensures that two different presentations cannot be associated to the same person. Notably, $A$ is allowed to provide the credentials $(cred_0, cred_1)$ as well as the issuer public keys, the disclosed attributes and the header of the presentation.

\begin{figure}
\tiny
\begin{algorithmic}

\State \boldsymbol{$Unlink^A$}$(1^\lambda, issuers)$
\Statex \makebox[0pt][l]{\rule{0.5\dimexpr\linewidth-\algorithmicindent}{0.4pt}}

    \State $b \overset{{\scriptscriptstyle \operatorname{R}}}{\gets} \{0,1\}$
    \State $((cred_0, cred_1), (ipk_0, ipk_1), disc\mhyphen attrs, header) \gets A()$
    \State $P_b \gets \mathsf{ABC.Prove}(ipk_b, cred_b, disc\mhyphen attrs)$
    \State $b' \gets A(P_b)$

    \If{$\mathsf{ABC.VerifyCred}(ipk_0, cred_0)$ \\
    $\land \mathsf{ABC.VerifyCred}(ipk_1, cred_1) \land b' = b$}

        \State $A$ wins the game
    \EndIf

\end{algorithmic}
\caption{Unlinkability game for an ABC scheme.}
\label{alg:abc-unlink}
\end{figure}

\subsection{Signature Scheme}
\label{app:sig}

The Existential Unforgeability under Chosen Message Attack $EUF-CMA^A$ security game for a signature scheme is shown in Figure \ref{alg:euf-cma}. $A$ is allowed to request signatures for arbitrary messages. To win the game the adversary $A$ must provide a forged signature for a message $m$, as long as she has not requested a signature for that message. The security game is extracted from \cite{suf-cma}, in which the authors prove that the signature scheme Ed25519 meets the requirement.

\begin{figure}
\tiny
\begin{algorithmic}

\State \boldsymbol{$EUF-CMA^A$}$(1^\lambda)$
\Statex \makebox[0pt][l]{\rule{0.5\dimexpr\linewidth-\algorithmicindent}{0.4pt}}

    \State $(spk, ssk) \gets \mathsf{S.GenKeyPair}(1^\lambda)$
    \State $sigs \gets []$
    \State $(m, \sigma) \gets A^Q(spk)$
    \Assert{ }
    \If{$\mathsf{S.Verify}(spk, m, \sigma) \land m \notin sigs$}
        \State $A$ wins the game
    \EndIf

\\

\State \boldsymbol{$Q_{Sign}$}$(m)$
\Statex \makebox[0pt][l]{\rule{0.5\dimexpr\linewidth-\algorithmicindent}{0.4pt}}

    \State $\sigma \gets \mathsf{S.Sign}(ssk, m)$
    \State $sigs \gets sigs + m$
    \State \Return{$\sigma$}

\end{algorithmic}
\caption{EUF-CMA game for a signature scheme.}
\label{alg:euf-cma}
\end{figure}

\begin{figure*}
\begin{multicols}{2}
\tiny
\begin{algorithmic}

\State \boldsymbol{$KInd$}$(1^\lambda, users)$
\Statex \makebox[0pt][l]{\rule{0.8\dimexpr\linewidth-\algorithmicindent}{0.4pt}}

    \State $state, epoch, chall \gets \{\}$
    \State $props, comms, I \gets []$
    \State $b \overset{{\scriptscriptstyle \operatorname{R}}}{\gets} \{0,1\}$
    \For{$id$ in $users$}
        \State $state[id] \gets \mathsf{Init}(1^\lambda, id)$
    \EndFor
    \State $b' \gets A^Q()$
    \If{$b' = b$}
        \State $A$ wins the game
    \EndIf

\State \boldsymbol{$Q_{Init}$}$(id)$
\Statex \makebox[0pt][l]{\rule{0.8\dimexpr\linewidth-\algorithmicindent}{0.4pt}}

    \Assert{ $state[id]$ does not exist}
    \State $state[id] \gets \mathsf{Init}(1^\lambda, id)$

\\

\State \boldsymbol{$Q_{Create}$}($id)$
\Statex \makebox[0pt][l]{\rule{0.8\dimexpr\linewidth-\algorithmicindent}{0.4pt}}

    \State $state[id] \gets \mathsf{Create}(state[id])$

\\

\State \boldsymbol{$Q_{Propose}$}$(id, id', prop\mhyphen type)$
\Statex \makebox[0pt][l]{\rule{0.8\dimexpr\linewidth-\algorithmicindent}{0.4pt}}

    \If{$prop\mhyphen type = \mathsf{add} \lor prop\mhyphen type = \mathsf{external}$}
        \State $JP \gets \mathsf{GenJP}(state[id'], prop\mhyphen type)$
    \EndIf
    \State $(state[id], prop) \gets \mathsf{Propose}(state[id], id', prop\mhyphen type, JP)$
    \State $props \gets props + prop$
    \State \Return{$(prop, JP)$}

\\

\State \boldsymbol{$Q_{Commit}$}$(id, (i_0, ..., i_n))$
\Statex \makebox[0pt][l]{\rule{0.8\dimexpr\linewidth-\algorithmicindent}{0.4pt}}

    \State $P_L \gets (props[i_0], ... props[i_n])$
    \State $(state[id], C) \gets \mathsf{Commit}(state[id], P_L)$
    \State $comms \gets comms + (epoch[id], C)$
    \State \Return{$C$}
    
\columnbreak

\State \boldsymbol{$Q_{Process}$}$(id, i_{C})$
\Statex \makebox[0pt][l]{\rule{0.8\dimexpr\linewidth-\algorithmicindent}{0.4pt}}

    \State $(ep, C) \gets comms[i_{C}]$
    \Assert $epoch[id] = ep$
    \State $(\gamma, ok) \gets \mathsf{Process}(state[id], C)$
    \If{$\lnot ok$} \State \Return{} \EndIf
    \State $state[id] \gets \gamma$
    \State $epoch[id] \gets epoch[id] + 1$
    \State $I[epoch[id]] \gets state[id].secret$

\\

%

\\

\State \boldsymbol{$Q_{Challenge}$}$(t)$
\Statex \makebox[0pt][l]{\rule{0.8\dimexpr\linewidth-\algorithmicindent}{0.4pt}}

    \Assert{ $I[t]$ exists $\land \lnot chall[t]$}
    \State $chall[t] \gets true$
    \State $k_0 \gets I[t]$
    \State $k_1 \overset{{\scriptscriptstyle \operatorname{R}}}{\gets} \mathcal{K}$
    \State \Return{$k_b$}

\\

\State \boldsymbol{$Q_{Reveal}$}$(t)$
\Statex \makebox[0pt][l]{\rule{0.8\dimexpr\linewidth-\algorithmicindent}{0.4pt}}

    \Assert{ $I[t]$ exists $\land \lnot chall[t]$}
    \State $chall[t] \gets true$
    \State \Return{$I[t]$}

\\

\State \boldsymbol{$Q_{Corrupt}$}$(id)$
\Statex \makebox[0pt][l]{\rule{0.8\dimexpr\linewidth-\algorithmicindent}{0.4pt}}

    \State \Return{$state[id]$}

\end{algorithmic}
\end{multicols}
\caption{Key Indistinguishability game for a CGKA scheme.}
\label{alg:cgka-kind}
\end{figure*}

\section{Security Proofs}

\subsection{Proof of Theorem \ref{th:ri}}
\label{app:pr_ri}


We will consider two different possibilities: either (1) the output contains a commit $C$ created in a previous epoch, or (2) otherwise. Clearly, these two cases are complementary.

\paragraph*{\textbf{(1) $A$'s output contains a commit $C$ created in a previous epoch}}

Let $ep_i$ and $ep_j$ be the epoch in which $C$ was created and the current epoch, respectively. The call to $\mathsf{S.Verify}$ in $\mathsf{Process}$ employs as input the challenge $\gamma.chal$, which is updated in every epoch using the commit that initiated it, as shown in auxiliary function $updateGroupChal(chal, C)$. In order for this signature to be verified correctly, the challenge of $ep_i$ must be the same as that of epoch $ep_j$. 

Let $B$ be an adversary attempting to find a collision for the hash function $H$, that acts as a wrapper for $A$. $B$ will simulate the Challenger for $A$'s game and execute all queries involved in it. $B$ also records all outputs of calls to $updateGroupChal$. Eventually, $A$ outputs $C$ and $B$, which executes $\mathsf{Process}$. If successful, then $B$ has found a collision between the challenge at the current epoch $ep_j$ and the challenge of epoch $ep_i$. As $H$ is collision-resistant, the probability of $C$ being accepted is negligible.

\paragraph*{\textbf{(2) $A$'s output does not contain a commit $C$ created in a previous epoch}}

In this scenario, $A$ forges a commit $C$ that was not legitimately created through an oracle call. The $\mathsf{Process}$ function in Figure~\ref{alg:def} verifies $C$'s signature with the signature public key $spk$ of the group member with ID $id$. The key pair of $id$ is generated in $\mathsf{Present}$. For our experiment, we consider the query $q_{create}$ to $\mathsf{Q_{Propose}}$  in which $id$'s key pair $(spk_{id}, ssk_{id})$ is created.

Let $B$ be an adversary playing a EUF-CMA game for the signature scheme $S$, that acts as a wrapper for $A$. $B$ will simulate the Challenger for $A$'s game and execute all queries involved in it. Upon receiving query $q_{create}$, $B$ will initiate its game to obtain $spk$ and set $id$'s public key to that value. $B$ will then proceed as normal, but by requesting a signature from the EUF-CMA Challenger to replace every call to $\mathsf{S.Sign}$ that involves $id$'s key pair. Then, upon receiving output $(id, C_{basic}, C_{reqs}, W, \sigma) \gets C$ from $A$, $B$ sends the message $((id, C_{basic}, C_{reqs}, W, state[id].chal), \sigma)$ to its Challenger. Clearly, if $A$ is able to produce a forgery, then $B$ would win its game. Since we assume the signature scheme is EUF-CMA secure, $B$'s probability of winning the game is negligible. Furthermore, it is possible for $\mathsf{Process}$ to reject $C$ even if the signature is successfully forged: thus, $A$'s probability of creating a forgery is lower than $B$'s probability of winning the EUF-CMA game, which is negligible. 


\subsection{Proof of Theorem \ref{th:unf}}
\label{app:pr_unf}


In order to add a new member to the AA-CGKA group, the commit $C$ in $A$'s output needs to contain a basic $\mathsf{add}$ proposal (as opposed to a \textit{requirement} $\mathsf{add}$ proposal). Clearly, neither $\mathsf{update}$ or $\mathsf{remove}$ proposals can result in any new members to the group. Whether $C$ was generated through an External Join or using the query $Q_{Commit}$, it must contain an $\mathsf{add}$ proposal. Let $PP_A = (P_A, JP_A)$ be the Presentation Package contained in the aforementioned proposal. We recall that all legitimately generated Presentation Packages are inserted in the list $PP_L$.

We now consider the situation in which $P_A$ has been created by $A$, whether it be by modifying some Presentation $P$ in $PP_L$ or by completely forging one. First, we rule out the situation in which $A$ uses one of the credentials contained in $Q_{creds}$ to create $P_A$, since we explicitly disallow this action in our definition of Unforgeability. We claim that $P_A$ will be rejected. Let $B$ be an adversary playing an Unforgeability game for the Attribute-Based Credential scheme $\mathsf{ABC}$, that acts as a wrapper for $A$. $B$ will simulate the Challenger for $A$'s game and execute all queries involved in it. In particular, $B$ will query its Challenger to issue new credentials, generate Presentations and expose credentials in $A$'s queries $Q_{Init}$, $Q_{Present}$ and $Q_{Expose}$, respectively. Upon $A$'s output $(id, C, k')$, B obtains $PP_A = (P_A, JP_A)$ from the proposal in $C$ and sends $P_A$ to its Challenger, along with the Presentation's $header$ and the issuer's identifier $i$. Since $B$'s chance of winning is negligible, $A$'s chance of producing a forged $P_A$ is negligible too. 

This means that, in order to be accepted, the Presentation Package in $A$'s output must contain a $P_A$ generated through a $Q_{Propose}$ oracle call, since any forgery by $A$ will be rejected with overwhelming probability. This includes the signature public key $spk$ that is part of $P_A$'s header. Let $PP_i \in PP_L$ be the Presentation Package from which $A$ obtains $P_A$, and $q_i$ the query in which $PP_i$ is generated.

Recall that a Join Package $JP_A = (JP_{TBS}, \sigma)$ is signed with $spk$. Since $spk$ has not been changed, we claim that the Join Package $JP_A$ could not have been modified either. Let $B$ be an adversary playing a EUF-CMA game for the signature scheme $S$, that acts as a wrapper for $A$. $B$ will simulate the Challenger for $A$'s game and execute all queries involved in it. Upon receiving query $q_i$, $B$ will initiate its game to obtain $spk$ and include it in the header of the Presentation $P_A$. $B$ will then proceed as normal, but by requesting a signature from the EUF-CMA Challenger to replace every call to $\mathsf{S.Sign}$ that involves $spk$. Then, upon receiving output $C$ from $A$, $B$ obtains the Presentation Package $PP_A = (P_A, (JP_{TBS}, \sigma))$ from it and sends the message $(JP_{TBS}, \sigma)$ to its Challenger. Clearly, if $A$ is able to produce a forgery, then $B$ would win its game. Since $B$'s chance of winning is negligible, $A$'s chance of producing a forged $P_A$ is negligible too. 

We have proven that $A$ cannot forge $P_A$ nor $JP_A$ (once $P_A$ is selected); therefore, $A$'s only chance of providing a valid Presentation Package is to completely copy some $PP_i \in PP_L$. Finally, we will prove that if $PP_A \in PP_L$, then $A$ would not be able to obtain $k$.

Let $G'$ be a game identical to $Unf^A$ as defined in Figure \ref{alg:sec-games}, but $A$'s first output is $(id, id', i)$, where $i$ is the index of an already generated Presentation Package. $C$ will obtain $PP_A \gets PP_L[i]$ and derive $prop\mhyphen type$ from the type of the Join Package contained in $PP_A$. As we have shown, $A$ could not insert a new user into the AA-CGKA group using a custom Presentation Package, so $A$'s probability of winning the Unforgeability game is the same as $G'$. We will call $q_i$ the query in which $PP[i]$ is created. Since Presentation Packages are created in $Q_{Propose}$ queries, $q_i$ will generate the $j$-th proposal stored in $props[j]$. 

Let $B$ be an adversary playing a Key Indistinguishability game for the CGKA scheme, that acts as a wrapper for $A$. $B$ will simulate the Challenger for $A$'s game $G'$ and execute all queries involved in it. Every call to a CGKA operation in $A$'s game will be replaced to an oracle call to $B$'s Challenger. Specifically, whenever $A$ calls $Q_{Propose}$, $B$ obtains the Join Package $JP_i$ by requesting its Challenger and then generates the Presentation Package $PP_i \gets (P_i, JP_i)$. Upon receiving $A$'s output $(id, id', i)$, $B$ will execute the query $C \gets Q_{Commit}(id, j)$ and then $Q_{Process}(id, i_C)$, where $i_C$ is the index of the commit generated. Then, $B$ will send $C$ to $A$. Upon receiving $k'$, $B$ will execute $Q_{Challenge}$ to obtain $k$. Since $B$'s chance of winning is negligible, $A$ will not be able to obtain any advantage in guessing $k$.

\subsection{Proof of Theorem \ref{th:unlink}}
\label{app:pr_unlink}


We assume that $\mathsf{CGKA.GenJP}$ outputs a fresh Join Package (i.e., that has not been previously used) and that it does not contain any information that could distinguish the users $id_0$ and $id_1$. This is consistent with our definition of a Join Package in Section \ref{sec:cgka}. Likewise, we assume that the key pair generated during the execution of $\mathsf{Present}$ employs a source of randomness such that it is impossible to correlate the generation of the key pair to any other action by the same user.

Let $B$ be an adversary playing an Unlinkability game for the Attribute-Based Credential scheme $ABC$, that acts as a wrapper for $A$. $B$ will simulate the Challenger for $A$'s game and execute all queries involved in it. When $A$ outputs $(id_0, id_1, GI)$, $B$ generates a key pair $(spk, ssk) \gets \mathsf{S.GenKeyPair}(1^\lambda)$ and a Join Package $JP \gets generateJoinPackage(1^\lambda)$ and send ($(cred_{id_0}, cred_{id_1})$, $(ipk_0, ipk_1)$, $GI.reqs$, $header = (GI.chal, spk)$) to its Challenger, where $(ipk_0, ipk_1)$ are the issuer public keys of $(cred_{id_0}, cred_{id_1})$, respectively. $B$'s Challenger will respond with $P_b$, and $B$ will send $PP_b = (P_b, JP)$ to $A$. $B$ will forward $A$'s response bit $b'$ to its Challenger. Since $B$'s probability of winning the $ABC$ Unlinkability game is negligible, $A$ will not be able to obtain any advantage in guessing $b$ and thus $A$'s probability of winning the games is also negligible.


\end{document}